\documentclass[twocolumn]{aastex63}

\usepackage{amsmath,amssymb}
\usepackage{graphicx}	
\usepackage{amsmath}	
\usepackage{amssymb}	
\usepackage{natbib}
\usepackage[flushleft]{threeparttable}
\let\tablenum\relax
\usepackage{siunitx}
\usepackage{enumitem}

\received{November 5, 2020}
\revised{January 8, 2021}
\accepted{January 17, 2021}

\submitjournal{ApJ}

\shorttitle{UV Spectral Slopes in the HFF}
\shortauthors{Bhatawdekar \& Conselice}

\graphicspath{{./}{figures/}}

\begin{document}

\title{UV Spectral-Slopes at $z=6-9$ in the Hubble Frontier Fields: Lack of Evidence for Unusual or Pop III Stellar Populations}

\email{Rachana.Bhatawdekar@esa.int}

\author[0000-0003-0883-2226]{Rachana Bhatawdekar}
\affiliation{European Space Agency, ESA/ESTEC, Keplerlaan 1, 2201 AZ Noordwijk, NL}
\affiliation{University of Nottingham, School of Physics \& Astronomy, Nottingham, NG7 2RD, UK}

\author[0000-0003-1949-7638]{Christopher Conselice}
\affiliation{University of Nottingham, School of Physics \& Astronomy, Nottingham, NG7 2RD, UK}
\affiliation{Jodrell Bank Centre for Astrophysics,University of Manchester, Oxford Road, Manchester, UK}

\begin{abstract}

We present new measurements of the UV spectral slope $\beta$ for galaxies at $z=6-9$ in the Frontier Fields cluster MACSJ0416.1-2403 and its parallel field, to an unprecedented level of low stellar mass. We fit synthetic stellar population models to the observed spectral energy distribution and calculate $\beta$ by fitting a power law to the best-fit spectrum. With this method, we report the derivation of rest-frame UV colours of galaxies for the Frontier Fields program extending out to $z=9$, probing magnitudes as faint as $M\mathrm{_{UV}=-13.5}$ at $z=6$. We find no significant correlation between $\beta$ and rest-frame UV magnitude $M_{1500}$ all redshifts, but we do find a strong correlation between $\beta$ and stellar mass with lower mass galaxies exhibiting bluer UV slopes. At $z=7$ the bluest median value of our sample is redder than previously reported values in the literature, whereas at $z=9$ our bluest data point has a median value of $\beta=-2.63_{-0.43}^{+0.52}$. Thus, we find no evidence for extreme stellar populations at $z>6$. We also observe a strong correlation between $\beta$ and SFR, such that galaxies with low SFRs exhibit bluer slopes. Additionally, there exists a star formation main sequence up to $z = 9$ with SFRs correlating with stellar mass. All of these relations show that $\beta$ values correlate with a process that drives both the overall SFR and stellar mass assembly. Furthermore, we observe no trend between $\beta$ and specific SFR, suggesting that $\beta$ is getting set by a global process driven by the scale of the galaxy.

\end{abstract}



\section{Introduction} \label{sec:intro}
At the highest redshifts, where only the rest-frame UV (dominated by emission from the most massive, young but short-lived hot stars) is currently accessible at high resolution imaging, one of the few characteristics of the physical properties of galaxies available is the UV colour, which is sensitive to star formation, dust and metallicity. Traditionally, changes in the UV colours are thought to be due to dust absorption. The inspection of dust in the early galaxies can be carried out through the measurement of the UV spectral slope, $\beta$, such that the UV spectral energy distribution has the form $f_{\lambda}\propto\lambda^{\beta}$ (e.g., \citealt{Calzetti1994}). It has also been found that these slopes are strongly correlated with dust extinction in galaxies at low redshift (e.g., \citealt{Meurer1995, Meurer1997, Meurer1999}) as well as at high redshift at $z\sim2$ (e.g., \citealt{Daddi2004, Reddy2012}). Therefore, these
values can be used to measure dust obscuration or extreme stellar populations at even higher redshifts.

The UV continuum slope $\beta$ has been extensively studied at high redshift ($z\geq2$) (e.g., \citealt{Bouwens2009, Bouwens2010, Finkelstein2010, Wilkins2011, Bouwens2012, Finkelstein2012, Dunlop2013, Oesch2013b, Rogers2014, Bouwens2014b, Kurczynski2014, Wilkins2016, Jiang2020}). For example, \citet{Bouwens2009} reported a strong evolution in the average values of $\beta$ from $-1.5$ at $z\sim2$ to  $-2.4 $ at $z\sim6$. In their study, they also found that lower luminosity galaxies appeared to be bluer than higher luminosity galaxies. Similarly, \citet{Wilkins2011} used single colours to measure the rest-frame UV colours of galaxies at $4.7<z<7.7$, and found that the mean UV continuum colours are approximately equal to zero (AB) for their highest redshift sample. At lower redshift, on the other hand, they find that the mean UV continuum colours of galaxies are redder, and furthermore find that galaxies with higher luminosities are also slightly redder on average. At $z\sim7$, \citet{Bouwens2010} measured $\beta$ for their sample of galaxies, finding that the very low luminosity galaxies exhibited UV continuum slopes as steep as $\beta=-3$. \citet{Finkelstein2010} also reported similar steep values of $\beta$ at $z\sim6-7$, albeit with larger uncertainties. More recently, \citet{Jiang2020} studied six luminous Ly$\alpha$ emitters (LAEs) at $z\sim6$ and reported very blue UV-continuum slopes in a range of $-3.4\leq\beta\leq-2.6$ at $M_{\mathrm{UV}}<-20$.

However, these findings are not without controversy. One example of this is that \citet{McLure2011} and \citet{Dunlop2012} find a variance-weighted mean value of $\beta\sim-2$ at $z\sim7$, and also find that $\beta$ shows no significant trend with either redshift or $M_{\mathrm{UV}}$. Similarly, using the imaging from UDF12 campaign with improved filter coverage and depth, \citet{Dunlop2013} calculate the UV colours of galaxies at redshifts $z>6.5$ and report similar redder colours with an average value of $\beta\sim-2$.

In another study, \citet{Bouwens2012} measured UV continuum slopes at $z\sim4-7$ and found that $\beta$ measurements for faint sources are likely to suffer large biases if the same passbands are used to select the sources as well as to measure $\beta$. They find that their high redshift  galaxies show a well-defined rest-frame UV colour--magnitude (CM) relationship that becomes systematically bluer toward fainter UV luminosities and that the dust extinction is zero at low luminosities and at high redshifts. Alternatively, \citet{Finkelstein2012} report no significant evolution of $\beta$ for galaxies at all luminosities within the GOODS-South and HUDF09 surveys at $z=4-8$. However, they suggest a significant correlation with stellar mass, such that more massive galaxies appeared redder. On the other hand, \citet{Bouwens2014b} measured a significant colour magnitude relation, with fainter galaxies displaying bluer slopes, such that the relation steepens at $z=4-8$. \citet{Wilkins2016} studied the rest-frame UV colours of four bright galaxies at $z\sim10$ in GOODS fields and a CLASH source behind MACS1149 cluster and report a measured $\beta$ of these candidates to be $-2.1\pm0.3$. More recently, \citet{Carvajal2020} stack the Lyman Break galaxies found in the Hubble Frontier Fields clusters with ALMA and report no trend between $\beta$ and redshift but a clear trend between stellar mass and $\beta$ similar to \citet{Finkelstein2012}.

Regardless of the varying results, now there is a broad agreement that atleast out to $z=6$, the values of $\beta\simeq-2$ are measurable for even the faintest galaxies detected with \textit{HST}, and at high redshift the dust extinction is significantly less than at lower redshift. However, the numbers of such faint sources are still quite small ($\sim30$ in the HUDF plus its two parallel fields), and are only found at $M_{\mathrm{UV}}<-17$. 

With the use of gravitational lensing, the Hubble Frontier Fields (HFF) has propelled the limits of current astronomical facilities until the James Webb Space Telescope (JWST) is launched by boosting the fluxes of the faint galaxies. Therefore, the HFF data can offer the first insights into the rest-frame UV colours of galaxies at $-17<\mathrm{M_{UV}<-13}$. This is particularly exciting as models predict that galaxies with $\beta\sim-3$ only exist at $\mathrm{M_{UV}>-17}$ \citep{Dunlop2013}. If we detect such blue slopes from the faint HFF galaxies, we will potentially discover the first evidence for unusual stellar populations: very low-metallicity, Pop III, extreme IMF or low dust star-formation in the early Universe.

With the subtraction method that we have developed in \citet{Bhatawdekar2019}, we have been able to probe magnitudes as faint as $M\mathrm{_{UV}=-13.5}$ in the HFF MACSJ0416.1-2403 cluster and its parallel field . In this paper, we use this data from the \textit{HST} imaging along with \textit{Spitzer} and ground-based VLT data of the MACSJ0416.1-2403 cluster and its parallel field to derive the measurement of rest-frame UV colours of galaxies out to $z=9$ for the Frontier Fields program, by investigating how $\beta$ evolves over the redshift range $z=6-9$ as well as UV luminosity in a wide magnitude range $-22<\mathrm{M_{UV}<-13}$.

The structure of this paper is as follows: In Section~\ref{sec:dataset} we describe the data used in this study. In Section~\ref{sec:methods} we detail the method developed to subtract the massive cluster galaxies on the critical line of the MACSJ0416.1-2403 cluster along with the construction of multiwavelength catalog, from 0.4 to 4.5$\mu$m, using \textit{HST}, \textit{Spitzer} and ground-based VLT data. Additionally, in this section, we outline the method used for photometric redshift calculation, the selection criteria used to build the sample of high-redshift galaxies at $z=6-9$ along with the SED fitting method we employed to calculate the value of the UV spectral slope $\beta$. In Section~\ref{sec:results}, we present our results by investigating the relationship between $\beta$ and $M_{1500}$, $\beta$ and stellar mass, $\beta$ and SFR, and finally stellar mass and SFR. Lastly, we present the conclusions of this work and summarize the results in Section~\ref{sec:summary}.
Throughout this paper, all magnitudes are in the AB system \citep{Oke1983}, a $\Lambda$CDM cosmology with $H_{0}$ = 70 km s$^{-1}$ Mpc$^{-1}$, $\Omega_{M} = 0.3$, and $\Omega_{\Lambda} = 0.7$ is assumed, and a \citet{chabrier2003} stellar initial mass function is used.

 \section{The Dataset} \label{sec:dataset}
\subsection{HST imaging}
As a part of the HFF program, MACSJ0416.1-2403, hereafter MACSJ0416, (RA: 04:16:08.9, Dec: -24:04:28.7) and its parallel field (RA: 04:16:33.1, Dec: -24:06:48.7) were observed between Jan 2014--Feb 2014 (Epoch 1) and July 2014--September 2014 (Epoch 2). In this work, we use the drizzled 60 mas pixel-scale v1.0 mosaics along with their RMS and weight maps provided on the HFF website \footnote[1]{http://www.stsci.edu/hst/campaigns/frontier-fields/FF-Data} by the Space Telescope Science Institute (STScI). We refer the reader to the STScI release documentation \footnote[2]{https://archive.stsci.edu/pub/hlsp/frontier/} for a detailed description of the data release.

The depths of these \textit{HST} images are calculated with 100s of $0\farcs2$ radius apertures placed in random positions in the images and estimating fluxes in them. In Table~\ref{tab:dataset_table} we specify the resulting 5$\sigma$ limiting magnitudes for the seven bands using our method. We note that the 5$\sigma$ limiting magnitudes of the cluster are brighter than the field, which is due to the fact that the cluster is overshadowed by the light of bright foreground galaxies \citep{Bhatawdekar2019}.

\begin{deluxetable}{cccc}
\tabletypesize{\small}
 \tablecaption{Description of the data we use in this study. The 5$\sigma$ depths are estimated by placing 100s of $0\farcs2$ radius apertures in random positions in \textit{HST} images, $0\farcs4$ radius apertures in HAWK-I image and $1\farcs4$ radius apertures in IRAC images. \label{tab:dataset_table}}
  \tablewidth{0pt}
  \tablehead{
    & \multicolumn{1}{c}{MACS0416 Cluster} & \multicolumn{1}{c}{MACS0416 Parallel} \\
    \hline
   \colhead{Filter} & \colhead{5$\sigma$ Depth} & \colhead{5$\sigma$ Depth} & \colhead{Instrument}
   }
  \startdata
         F435W &28.87 &28.91 &ACS\\
		 F606W &28.95 &29.01 &ACS\\
		 F814W &29.35 &29.40 &ACS\\
		 F105W &29.22 &29.30 &WFC3\\
		 F125W &28.95 &28.02 &WFC3\\
		 F140W &28.85 &28.93 &WFC3\\
		 F160W &28.65 &28.75  &WFC3\\
		 Hawk-I $K_{s}$  &26.25 &26.35  &HAWK-I\\
		 IRAC 3.6 &25.10 &25.16 &IRAC\\
		 IRAC 4.5 &25.13 &25.20 &IRAC\\
  \enddata
\end{deluxetable}

\subsection{VLT imaging} \label{sec:vltdata}
Traditionally, to study the stellar masses and stellar populations of $z\gtrsim6$ galaxies, the best approach is to use \textit{Spitzer}/IRAC data at $> 3 \mu$m in combination with \textit{HST} imaging and $K_{s}$ band data. To fully exploit the poorer resolution \textit{Spitzer}/IRAC data and to put tight constraints on redshift measurements, we therefore introduce the longer wavelength $K_{s}$ band data at 2.2 $\mu$m, which helps fill the gap between the 1.6$\mu$m (F160W) band and the 3.6$\mu$m (IRAC) channel. For this, the fully reduced $K_{s}$ band images made available through the Phase 3 infrastructure of the ESO Science Archive Facility (ESO program 092.A-0472, P.I. Brammer) are used in this work. For a detailed description of the observations we refer the reader to \citet{Brammer2016}.

The depth of the image is measured by placing 100s of $0\farcs4$ radius apertures in random positions in the image and estimating fluxes in them, similar to the \textit{HST} bands. In Table~\ref{tab:dataset_table} we list the 5$\sigma$ limiting magnitudes for the $K_{s}$ band. 

\subsection{Spitzer imaging} \label{sec:spitzerdata}

The \textit{Spitzer} Space Telescope has devoted $\sim$ 1000 hours of Director's Discretionary time to observe the Frontier Fields at 3.6 $\mu$m and 4.5 $\mu$m. As the Balmer break is vital in the estimation of galaxy stellar mass, and is observed at wavelengths beyond 2.4 $\mu$m at $z>5$, we include the final reduced mosaics of \textit{Spitzer} data made available through the IRSA website \footnote[3]{http://irsa.ipac.caltech.edu/data/SPITZER/Frontier/} (Program ID 90258, P.I T. Soifer). In addition to getting robust stellar mass measurements, IRAC data is also crucial to put better constraints on redshift estimates.

Analogous to \textit{HST} and the $K_{s}$ band data, the depth of \textit{Spitzer} images is also calculated by placing 100s of $1\farcs4$ apertures in random positions in the images and measuring fluxes in them. We list the 5$\sigma$ limiting magnitudes for both the channels in Table~\ref{tab:dataset_table}.

\section{Methods}\label{sec:methods}
\subsection{Subtraction of massive galaxies}
While the power of gravitational lensing of massive clusters enables us to find and study the faintest galaxies in the Universe by providing a magnified boost to light from the background galaxies, in practice the light of the massive foreground galaxies in clusters makes this process difficult. Thus, it is essential to model and subtract these massive galaxies before doing any further analysis. 

\begin{figure*}
\centering
\begin{minipage}{0.49\textwidth}
\centering
\includegraphics[width=1\textwidth, height=0.33\textheight]{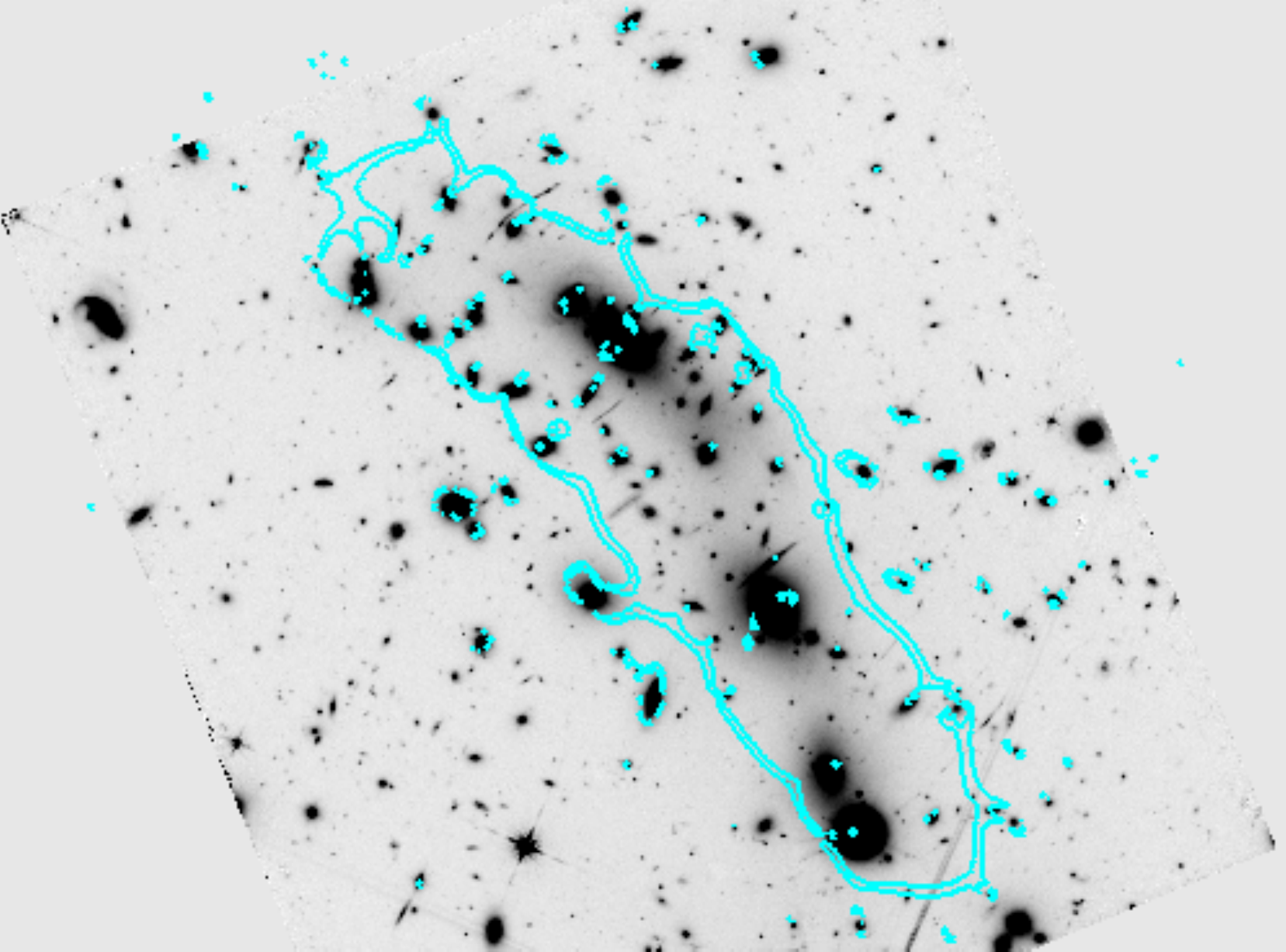}
\end{minipage}
\begin{minipage}{0.49\textwidth}
\centering
\includegraphics[width=1\textwidth, height=0.33\textheight]{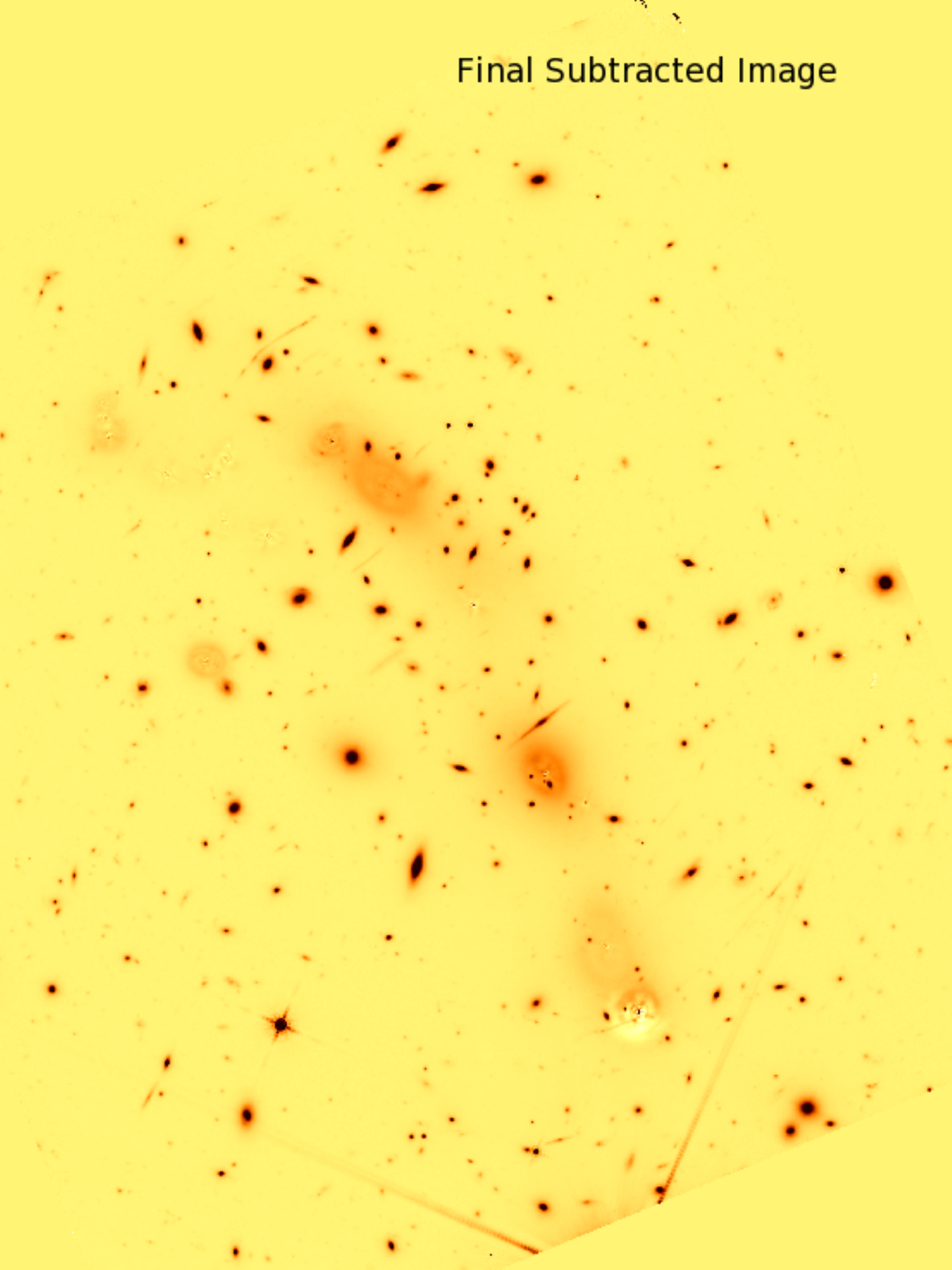}
\end{minipage}
\caption{F160W image of the MACSJ0416 cluster. The critical lines at $z=9$ from the CATS model \citep{Jauzac2014} are shown in cyan (left). The final subtracted image of the MACSJ0416 cluster (right).}
\label{fig:critical_lines}
\end{figure*}

For this, we developed a technique that accurately removes the most massive foreground galaxies from the critical lines of the MACSJ0416 cluster (as shown in Fig.~\ref{fig:critical_lines}) allowing for a deeper detection of the faint background galaxies in the H-band (reddest detection band). This is described in detail in \citet{Bhatawdekar2019} but, briefly, we use an iterative process in which we divide the image into small regions with the target bright galaxy in the center along with the small neighbouring galaxies. We then use GALAPAGOS \citep{Barden2012} and GALFIT \citep{Peng2002} on those rectangular regions to model the small galaxies first before trying to model the massive galaxies. The small galaxies are fitted one at a time with one or more S\'ersic \citep{Sersic1968} components until a reasonable residual is obtained. This process is carried out until all the small galaxies are fitted accurately and we are left with the central bright galaxy. The process is then repeated on the central galaxy until we get a good fit, after which we do a massive simultaneous fitting with all the neighbours. An image is then created with all these objects subtracted from the original image before moving on to fit the next rectangular patch to repeat the process in iteration. We refer the reader to \citet{Bhatawdekar2019} for complete details of our subtraction procedure.

\subsection{Multiwavelength photometry, photometric redshifts, and stellar mass} \label{sec:properties}
After subtracting the massive galaxies from the H-band of the MACSJ0416 cluster, we construct a multiwavelength photometry catalog from 0.4 to 4.5$\mu$m. The details of how we obtain the photometric measurements are explained in \citet{Bhatawdekar2019} but, briefly, for \textit{HST} images, SExtractor \citep{Bertin1996} is used in dual image mode with the subtracted H-band as the detection image and by using the same apertures to execute photometry on the rest of the bands for the MACSJ0416 cluster. The same method is applied on the parallel field, except for the subtraction process applied on the cluster. 

For photometry on the $K_{s}$ band and \textit{Spitzer} imaging, we use the T-PHOT code \citep{Merlin2015}. We refer the reader to \citet{Bhatawdekar2019} for the details of photometric measurements with T-PHOT. Finally, we construct a catalog of fluxes combining the photometry of \textit{HST}, VLT and \textit{Spitzer} imaging for the MACSJ0416 cluster and the parallel field.

After the photometry catalog is constructed, we estimate the photometric redshifts for our multiwavelength photometric measurements using  EAZY \citep{Brammer2008}. \citet{Dahlen2013} have shown that no specific set of template SEDs or code produce considerably better estimates of photometric redshifts compared to others. But they find that the codes that result in lowest scatter and outlier fraction typically use a training sample to optimize the photometric redshifts. Therefore, in our work, to calibrate the photometric redshifts, we use the CANDELS dataset \citep{Guo2013} trimmed down to only the filters we have and use those to optimise the EAZY parameters, which is then applied to our data. We then use EAZY and build a sample of galaxies in the redshift range  $5.5\leq z\leq9.5$ using a sample selection criteria described in detail in \citet{Bhatawdekar2019}. Briefly, we use the full redshift probability distribution function (PDF) ($P(z)\propto\exp(-\chi^{2}/2$)), using the $\chi^{2}$ distribution from EAZY and form galaxy samples in four redshift bins centered at z $\sim$ 6, 7, 8, and 9 with $\triangle z$ =  1, for both the MACS0416 cluster and the parallel field, by applying a set of additional selection criteria as follows:

\begin{equation}
\int_{z_{\mathrm{s}}-0.5}^{z_{\mathrm{s}}+0.5}p(z)dz>0.4
\end{equation}

\begin{equation}
\int_{z_{\mathrm{p}}-0.5}^{z_{\mathrm{p}}+0.5}p(z)dz>0.6
\end{equation}

\begin{equation}
(\chi_{\mathrm{min}}^{2}/N_{\mathrm{filters}}-1)<3,
\end{equation}

\noindent where $z_{s}=$ 6, 7, 8 and 9 for the respective bins and $z_{p}$ is the primary redshift peak.

The first criterion ensures that a significant area of the probability distribution function lies within the redshift range of our interest. With the second criterion we ensure that at least 60 per cent of the PDF lies near the peak of the distribution, such that the high-redshift solution is the dominant one. The third criterion is to make sure that EAZY provides a reasonable fit. Additionally, we place a S/N cut such that S/N$(J_{125})>3.5$ and S/N$(H_{160})>5$.

Following the above selection criteria, we also visually inspect each object to exclude potential contaminants such as stars, stellar diffraction spikes, sources at the edge of the images, sources with flagged photometry etc. Our final sample contains 134 galaxies: 92 at $z\sim6$, 24 at $z\sim7$, 10 at $z\sim8$ and 8 at $z\sim9$.

We also compare our estimated photometric redshifts with available spectroscopic redshifts, using the published redshift catalog of the MACS0416 cluster that combines the VIMOS CLASH-VLT campaign \citep{Balestra2016} and the MUSE spectroscopic study \citet{Caminha2017}. Following \citet{Dahlen2013}, we compute $\Delta z/(1+z_{\mathrm{spec}})$ (where $\Delta z=(z_{\mathrm{spec}-}z_{\mathrm{phot}})$) and find that our redshift accuracy is quite good, with a scatter of $\sigma_{\Delta z/(1+z_{\mathrm{spec}})}=0.041$. This is explained in detail in \citet{Bhatawdekar2019}.

Finally, we measure the stellar masses and rest-frame magnitudes for our sample by using a stellar population fitting technique as described in detail in \citet{Bhatawdekar2019}. Briefly, the stellar masses and rest-frame magnitudes are calculated using a custom template-fitting routine \textbf{SMpy}\footnote[6]{https://github.com/dunkenj/smpy} \citep{Duncan2014}. We first use the single stellar population models of \citet{Bruzual2003} to construct synthetic SEDs for a user-defined combination of parameters such as metallicity, age and star formation history (SFH), adopting a Chabrier IMF \citep{chabrier2003}. We allow the ages to vary from 5 Myr up to the age of the Universe at the redshift step being fit. We then vary dust attenuation in the range $0\leq A_{v}\leq2$, and metallicities of 0.02, 0.2 and 1 Z$_{\odot}$ are used. 

Typically, while modeling SEDs we need to assume some parameterization of the SFH, and this can introduce systematics. Non-parametric SFH reconstruction techniques therefore offer the best prospects of delivering less biased results, but they are computationally expensive and also require data of very high quality with a wide wavelength coverage, high S/N ratio ($>50$ angstrom) and a high spectral resolution to retrieve complex SFHs accurately (e.g., \citealt{Ocvirk2006,Tojeiro2007}). Due to these constraints, we choose to characterize the SFH with the widely adopted $\tau$-model. Previous work (e.g., \citealt{Lee2009}) have shown that single component exponentially decreasing SFHs result in significantly underestimated SFRs while \citet{Maraston2010} concluded that models with exponentially rising SFH provide better fits to observed SEDs of high redshift galaxies and also produce SFRs consistent with other methods. The conclusions from these analyses, thus, are that the model SFH must be adequately diverse to allow for a broad range in SFH forms. Therefore, while we choose to use the $\tau$-model in this work, we incorporate a wide range of SFHs in our work by considering various SFH histories; exponentially decreasing, increasing as well as constant SFR. To do this, we use the  universally assumed parametrization of the SFH (SFR $\propto e^{-t/\tau}$) with $\tau=0.05,0.25,0.5,1,2.5,5,10,-0.25,-0.5,-1,-2.5,-5,-10$ and $1000$ (constant SFR) Gyr. Here, negative $\tau$ values are used to represent exponentially increasing histories. 

While recovering complex SFHs with non-parametric methods is out of the scope of this study, it is worth considering whether our choice of $\tau$-model with a wide range of SFHs has any influence on the derived stellar masses as well as SFRs. It is well known that the outshining effect results in underestimated stellar mass when single-component SFH models are used \citep{Papovich2001}. However, as pointed out in \citet{Madau2014}, even with complex SFHs, the outshining effect will tend to underestimate the galaxy stellar mass. Nevertheless, at very high redshifts such as the redshifts probed in this work, the uncertainties due to SFH are reduced simply because the Universe is too young, such that even the oldest stars in galaxies must be younger than that. This therefore sets a limit on the M/L ratio for older stars and their possible contribution to the stellar mass of a galaxy. Additionally, \citet{Lee2010} have also reported that stellar mass appears to be the most robustly measured parameter irrespective of the assumptions in the SFH. Due to these reasons, we conclude that our choice of $\tau$-model with a wide range of SFHs does not influence the stellar mass estimates in this work. Furthermore, as will be shown in Section~\ref{sec:betasfr}, we find that models with a wide range of SFHs used in our work produce SFRs that are in good agreement with SFRs estimated with other methods, apart from a few galaxies with very high SFRs (SFR $>100\mathrm{M_{\odot}yr^{-1}}$). 

While analysing a large sample of Lyman break galaxies \citet{Schaerer2012} reported that the majority of objects were better fit with SEDs that accounted for nebular emission. Previous other studies have also found that including nebular emission lines in SED fitting result in considerably younger ages and lower masses (e.g., \citealt{Schaerer2009, Schaerer2010, Ono2010,McLure2011, Duncan2014}). We therefore choose to apply nebular emission lines on the model SEDs in this work. We refer the reader to \citet{Duncan2014} for the detailed description of the method employed to include nebular emission lines. We then apply dust extinction using the law described by \citet{Calzetti2000}.

Each model SED is then redshifted in the range $0\leq z\leq11$ in steps of $\triangle z=0.02$ and attenuation by neutral hydrogen is applied according to \citet{Madau1995}. Lastly, each model spectrum is convolved through the photometric filters and the following SED grid is fitted to the photometry. For each model, the absolute magnitude at 1500 angstrom is estimated by fitting a 100 \AA -wide top-hat filter centered on 1500 \AA. With a Bayesian-like approach, the model SEDs are then fitted to the observed photometry, which results in a likelihood distribution of stellar mass and rest-frame magnitudes. Additionally, we calculate the SFRs from the SED fitting code by obtaining the SFR from the best-fitting template for each galaxy.

Finally, for the Frontier Fields program, lensing models were produced by seven independent teams\footnote[7]{https://archive.stsci.edu/prepds/frontier/lensmodels/} and we make use of all the models to calculate the median magnification value with which we demagnify the rest-frame magnitudes, stellar masses and star formation rates of our sample in the MACS0416 cluster.

\subsection{Calculating UV spectral slope $\beta$}\label{sec:betacal}
Once we have the multiwavelength photometry catalog and the sample of high redshift galaxies in the redshift range  $5.5\leq z\leq9.5$, the next step is to estimate the UV spectral slope $\beta$. The UV continuum, approximately between 1250 angstrom and 2600 angstrom, was first parametrized by \citet{Calzetti1994} as a power law of the form $f_{\lambda}\propto\lambda^{\beta}$ to study the effects of dust extinction in starbursting galaxies. The UV spectral slope $\beta$ was designed to be measured from spectra in the specified wavelength windows defined by \citet{Calzetti1994}. However, because continuum spectroscopy is exceptionally challenging at high redshift, $\beta$ is usually determined by three main methods, each with its own advantages and disadvantages.

The first method is by fitting a power-law to all the available photometry redward of the Lyman break to measure $\beta$ (e.g., \citealt{Bouwens2014b}). A direct fit to the photometry can hence allow for measurements of $\beta$ for faint galaxies that are close to the detection limit. Translating the $f_{\lambda}\propto\lambda^{\beta}$ relation into magnitude units gives a linear relationship between $\beta$ and colours. Therefore, another way of estimating $\beta$ is via a single colour (e.g., \citealt{Dunlop2012, Hathi2008}). The third method of calculating  $\beta$ is by performing SED fitting (e.g., \citealt{Finkelstein2012}). In this method, single stellar population models are first constructed by varying parameters such as age, metallicity, star formation history and dust. The best-fit model is then found via  $\chi^{2}$ minimization, and the value of $\beta$ is measured directly from this best-fit spectrum by fitting a power law to the spectrum using the 10 wavelength windows specified by \citet{Calzetti1994}. A primary difficulty in determining $\beta$ for faint galaxy populations results from the noisy photometry that make inferring colors between bands particularly fraught. This method of fitting SED models to the data and using the models to infer $\beta$ has the potential to overcome this issue by using additional information from the rest-frame optical to model the overall spectrum of each galaxy, or at least to ameliorate the effects of filter-to-filter scatter on the inferred beta. The advantage of this method is that because all the available photometric bands are used, this should, in principle, yield more robust estimates of $\beta$. The disadvantage, however, is that because this method is based on synthetic models, we are confined to a limited range of $\beta$ values allowed by the models. Therefore, this method may not be suitable to the population of objects close to the epoch of first star formation that may have unique spectral features. 

In our work, because we have multiple photometric bands, and hence multiple rest-frame UV colours redward of the Lyman break, we choose to use the SED fitting method and follow the procedure described in \citet{Finkelstein2012}, who show with the help of simulations that SED fitting is a superior choice over the other two methods. 

To measure $\beta$, we perform SED fitting on our sample of high redshift galaxies at $z=6-9$ to find the best fitting synthetic stellar population models of \citet{Bruzual2003}.  With the help of synthetic SEDs constructed from the single stellar population models of \citet{Bruzual2003} explained Section~\ref{sec:properties}, we find the best-fit model via $\chi^{2}$ minimization, and measure the value of $\beta$ directly from this best-fit spectrum by fitting a power law to the spectrum using the 10 wavelength windows specified by \citet{Calzetti1994}. 

In order to estimate the uncertainty on $\beta$, we perform a Monte Carlo analysis in which we perturb the observed flux of each source by randomly choosing a point from a Gaussian distribution, the standard deviation of which is the 1$\sigma$ uncertainty on the flux in any given filter. The UV slope $\beta$ for each source is then estimated with the simulated photometry by deriving a best-fit model. This process is repeated 500 times and the final uncertainty is then taken as the standard deviation of the distribution of these 500 values. Whilst we do a careful subtraction of the foreground cluster galaxies such that there are no over-subtracted residuals (as seen Fig~\ref{fig:critical_lines}), which could lead to noisy photometry, this step ensures that the photometric errors are taken into account while estimating the uncertainty on $\beta$.

As identified by \citet{Finkelstein2012}, this SED fitting method has a disadvantage in that we are confined to the limited range of $\beta$ values that are allowed in the chosen model. Therefore, the UV colours of galaxies bluer than the allowed $\beta$ values in a particular model will not be recovered precisely. For our set of models the bluest value is $\beta=-3.1$, but since we are not finding extreme blue colours, the bluest value we find in our sample is $\beta=-2.82_{-0.24}^{+0.65}$, we are not affected by this limitation as we are not approaching our limit of bluest parameter space. Similarly, we explore whether the choice of a particular IMF could have any impact on the derived values of $\beta$. To do this, we estimate our $\beta$ values also with a \citet{Salpeter1955} IMF in our SED fitting code and find that the choice of IMFs has no affect on the measured values of $\beta$ as long as the model fits the data well. This is ensured by our third selection criteria (See Section~\ref{sec:properties}), with which we make sure that EAZY provides a reasonable fit to the data. As pointed out in \citet{Finkelstein2012}, this will be true for the choice of models also, as presuming that the model is a good fit to the data, two models with different ages or SFHs will have comparable UV slopes. These results are also consistent with \citet{Jerabkova2017} in which they compute UV slope for various IMFs and show that the $\beta$ values are dominated by the age of stellar population, and apart from very early phases ($<$10 Myr) the difference is overall very small irrespective of the choice of IMFs.

\begin{figure*}
\centering
\begin{minipage}{0.45\textwidth}
\centering
\includegraphics[width=1\textwidth, height=0.3\textheight]{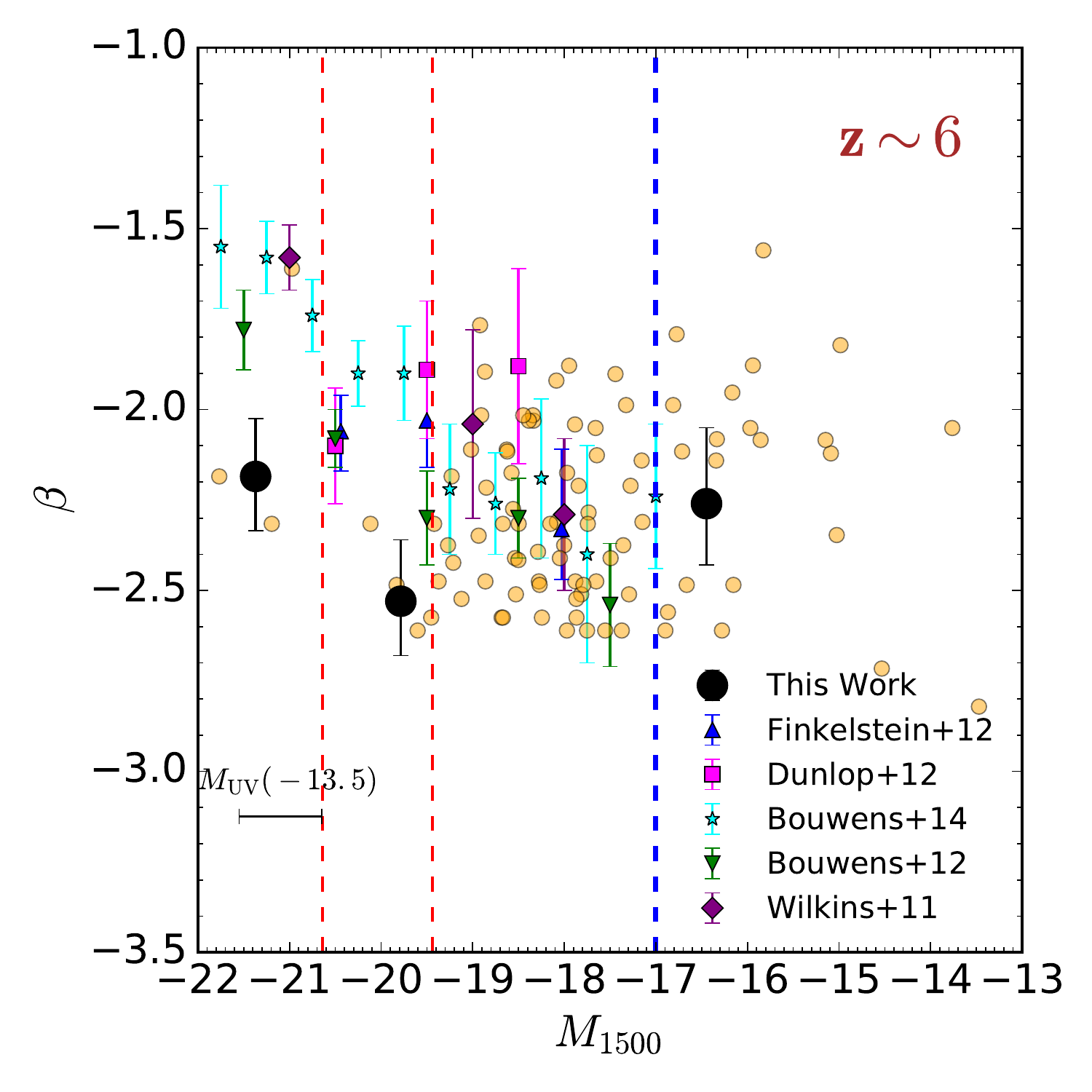}
\end{minipage}
\begin{minipage}{0.45\textwidth}
\centering
\includegraphics[width=1\textwidth, height=0.3\textheight]{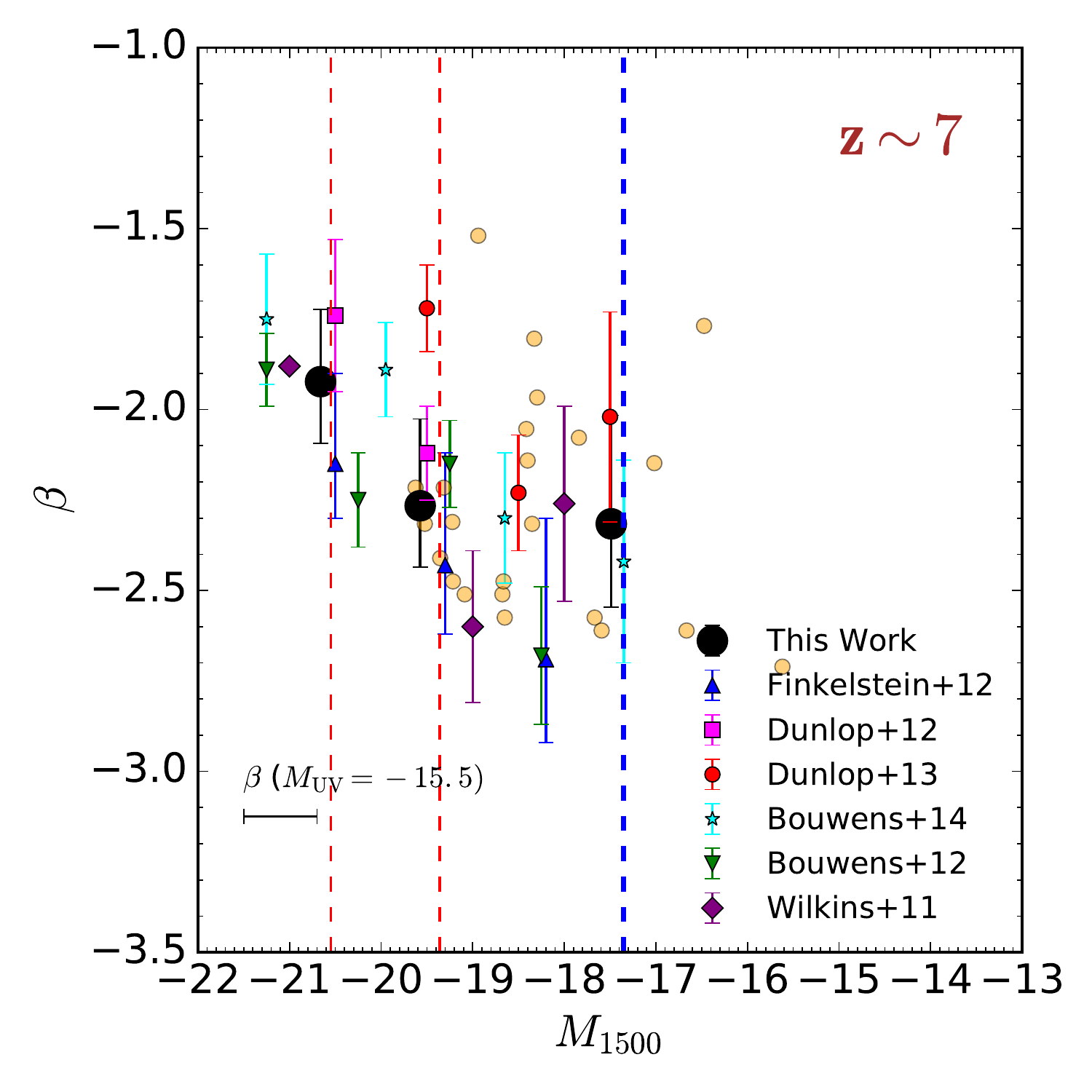}
\end{minipage}
\begin{minipage}{0.45\textwidth}
\centering
\includegraphics[width=1\textwidth, height=0.3\textheight]{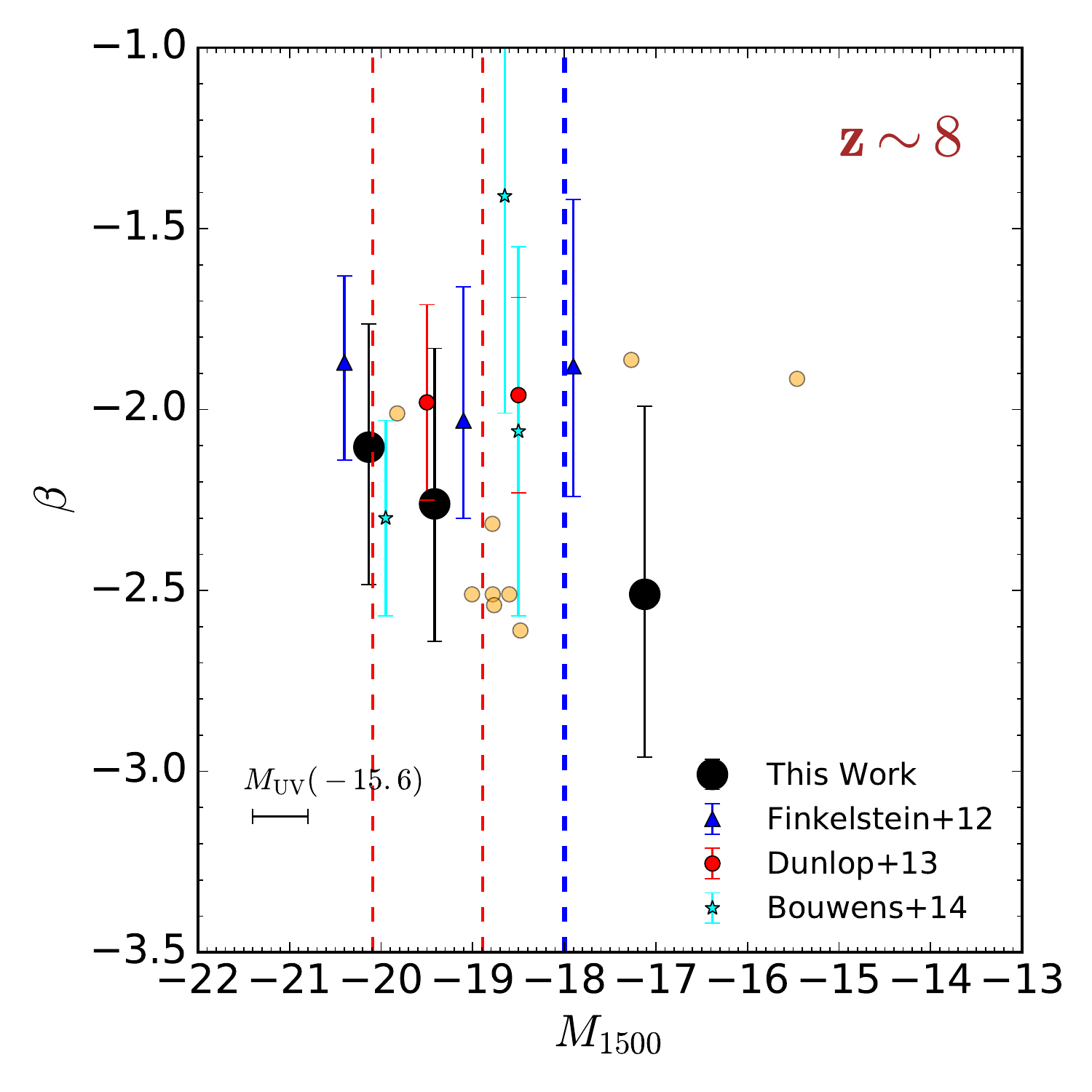}
\end{minipage}
\begin{minipage}{0.45\textwidth}
\centering
\includegraphics[width=1\textwidth, height=0.3\textheight]{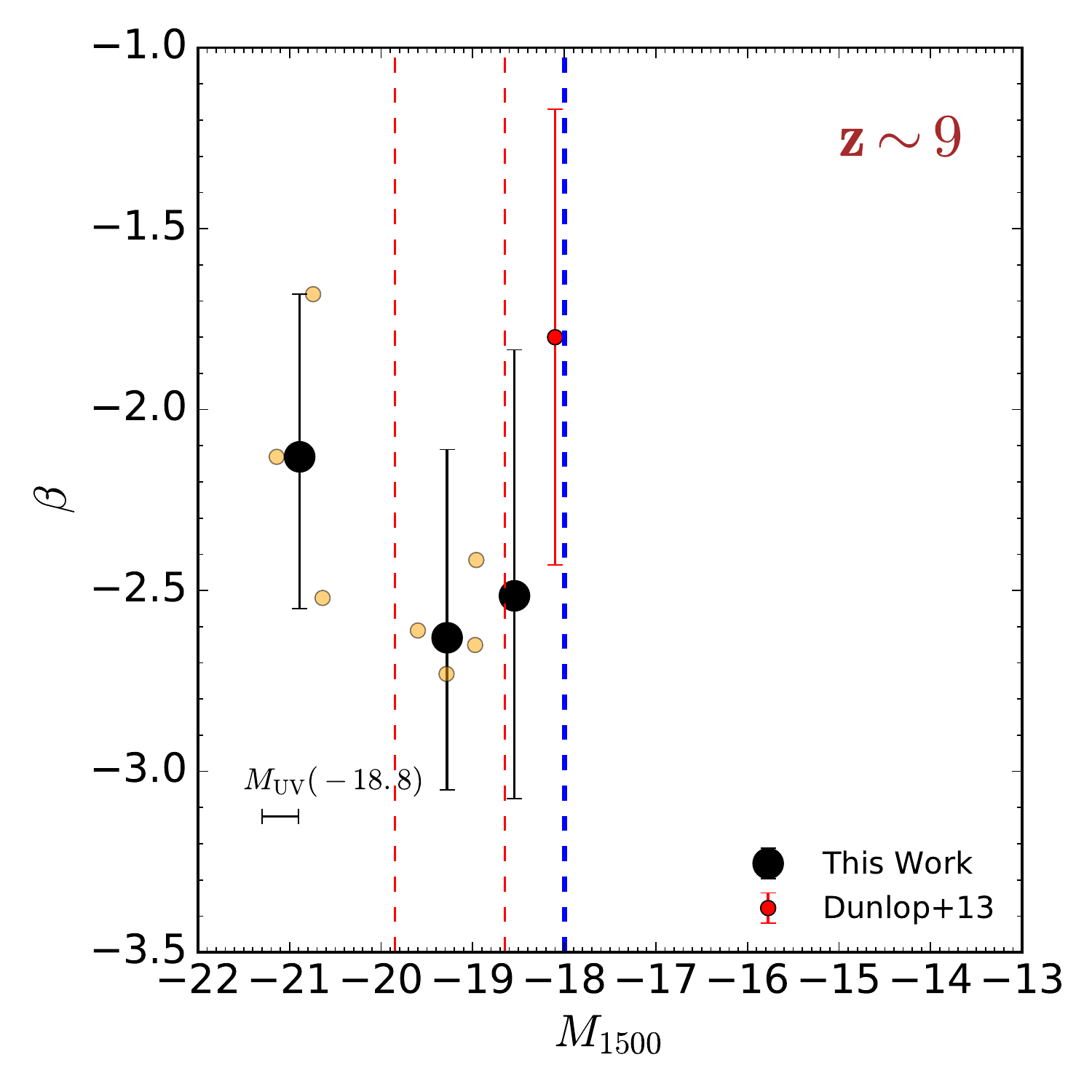}
\end{minipage}
\caption{The estimated UV slope $\beta$ vs absolute magnitude at 1500 Angstrom at $z=6-9$. The filled yellow circles show the results for individual galaxies, whereas the black circles
show the median value of $\beta$ measured in three different bins separated by the red dashed lines, denoting the 0.25 $L^{*}$ and 0.75 $L^{*}$ values, similar to \citet{Finkelstein2012}, with the uncertainties being the errors on the median, estimated with bootstrap Monte Carlo simulations. The blue triangles are the median values of $\beta$ from \citet{Finkelstein2012}, green triangles are the biweighted mean UV continuum slopes from \citet{Bouwens2012}, cyan stars are biweighted mean $\beta$ values from \citet{Bouwens2014b}, purple diamonds are the mean $\beta$ values from \citet{Wilkins2011}, magenta squares are the average $\beta$ values from \citet{Dunlop2012} and red circles are the derived mean $\beta$ values from two-band colours from \citet{Dunlop2013} for comparison. The representative error bar at the bottom left of each plot shows the measured error on the faintest galaxies in each bin due to lensing. The dashed blue line shows the magnitude limit of previous studies at each redshift.}
\label{fig:uv_beta}
\end{figure*}

Fig.~\ref{fig:uv_beta} shows the estimated values of the UV slope $\beta$ for each of our high redshift sources at $z=6-9$ as a function of $M_{\mathrm{1500}}$. As the errors on individual measurements of $\beta$ are large, notably at higher redshifts and fainter magnitudes (See Section~\ref{sec:testingmethod}), we follow \citet{Finkelstein2012} and calculate the median values of $\beta$ in three different bins, separated by 25 per cent and 75 percent of the characteristic magnitude $L^{*}$ values, using our derived luminosity functions in \citet{Bhatawdekar2019}. Regardless of our choice of bins, we do test whether a choice of different bin size could affect our results and we find that our results are consistent irrespective of the choice of bins.

We perform bootstrap Monte Carlo simulations to calculate the errors on our median values of $\beta$ by accounting for Poisson noise as well as photometric noise. For this, we take our original sample of galaxies in each luminosity bin and create new simulated samples from them to account for the Possion noise. We do this by first random sampling with replacement from a normal distribution. To account for the photometric error, we then take this modified sample and remeasure $\beta$, again by random sampling from a normal distribution and by taking into account the photometric uncertainty in $\beta$ for each galaxy. This process is repeated for $10^{4}$ times for each luminosity and redshift bin as well as for all the galaxies. The final uncertainty is then taken as the standard deviation of these new simulated values of $\beta$.

In Fig.~\ref{fig:uv_beta} we show these median values of $\beta$ with the associated uncertainty in black circles. Similarly, in Table~\ref{tab:uv_beta_table} we list the median values of $\beta$ for all the galaxies in each redshift bin, along with the median values in the three different bins. Additionally, we specify the median $M_{\mathrm{1500}}$ values at each redshift.
 
\begin{deluxetable*}{ccccccc}
\tabletypesize{\small}
\tablecaption{Median values of the UV spectral slope $\beta$. Column (1) lists the redshifts, Column (2) lists the median $\beta$ values for all galaxies in each redshift bin.  Column (3), Column (4) and Column (5) list the
median values of $\beta$ in three different bins, separated by 25 per cent and 75 percent of the characteristic magnitude $L^{*}$ values, using our derived luminosity functions in \citet{Bhatawdekar2019}, Column (6) lists the values of fitted slope and its associated uncertainty, whereas Column (7) lists the median $M_{\mathrm{1500}}$ values for all galaxies in each redshift bin.}
\label{tab:uv_beta_table}
\tablewidth{0pt}
\tablehead{
\colhead{$z$} & \colhead{Median $\beta$} & \colhead{Median $\beta$} & \colhead{Median $\beta$} & \colhead{Median $\beta$} & \colhead{$\beta$ - $M_{\mathrm{1500}}$ slope} & \colhead{Median $M_{\mathrm{1500}}$} \\[-0.1cm]
\colhead{$ $} & \colhead{All Galaxies} & \colhead{$L>0.75L^{*}$} & \colhead{$0.25L*<L<0.75L*$} & \colhead{$L<0.25L^{*}$} & \colhead{All Galaxies} & \colhead{}
}
\startdata
	 6 & $-2.22_{-0.12}^{+0.08}$ & $-2.18_{-0.15}^{+0.16}$ & $-2.53_{-0.15}^{+0.17}$ & $-2.26_{-0.17}^{+0.21}$ & $-0.01\pm0.06$ &$-17.96$\\
		 7 & $-2.31_{-0.16}^{+0.23}$ & $-1.92_{-0.17}^{+0.20}$ & $-2.26_{-0.20}^{+0.24}$ & $-2.32_{-0.23}^{+0.30}$ & $-0.11\pm0.07$ & $-18.53$ \\
		 8 & $-2.41_{-0.22}^{+0.26}$ & $-2.10_{-0.38}^{+0.34}$ & $-2.26_{-0.42}^{+0.43}$ & $-2.51_{-0.45}^{+0.52}$ & $-0.13\pm0.07$ & $-18.77$\\
		 9 & $-2.52_{-0.20}^{+0.32}$ & $-2.13_{-0.42}^{+0.45}$ & $-2.63_{-0.43}^{+0.52}$ & $-2.51_{-0.56}^{+0.68}$  & $-0.19\pm0.11$ & $-19.44$\\
\enddata
\end{deluxetable*}

\subsection{Goodness of method}\label{sec:testingmethod}
In order to test the level of quality of our method estimating $\beta$, we construct a simulated catalog of high redshift galaxies using the Theoretical Astrophysical Observatory (TAO) \citep{Bernyk2016}. For this, we use the existing CANDELS mock light cone on the TAO from redshift $z=0$ to $z=9$ and create SEDs from the single stellar populations of \citet{Bruzual2003} with the initial mass function of \citet{chabrier2003}, similar to our selection efficiency method in \citet{Bhatawdekar2019}. We then apply dust with the dust model of \citet{Calzetti2000} and make the final catalog with an H-band distribution of magnitudes in the range $21<H_{160}<35$. The magnitudes in the remaining filters are then deduced from the range of H-band magnitudes. We then directly fit this spectrum to get the known input value of $\beta$.

To generate errors for the simulated photometry catalog, we bin the object fluxes from our real catalog and compute the mean and standard deviation of the error on fluxes in those bins. This provides us with a Gaussian distribution of the errors. We then simulate the photometric errors in the fake catalog by choosing random errors each time from the Gaussian distribution. This simulated catalog is then run through EAZY to get the photometric redshifts. 

\begin{figure*}
\centering
\begin{minipage}{0.45\textwidth}
\centering
\includegraphics[width=1\textwidth, height=0.3\textheight]{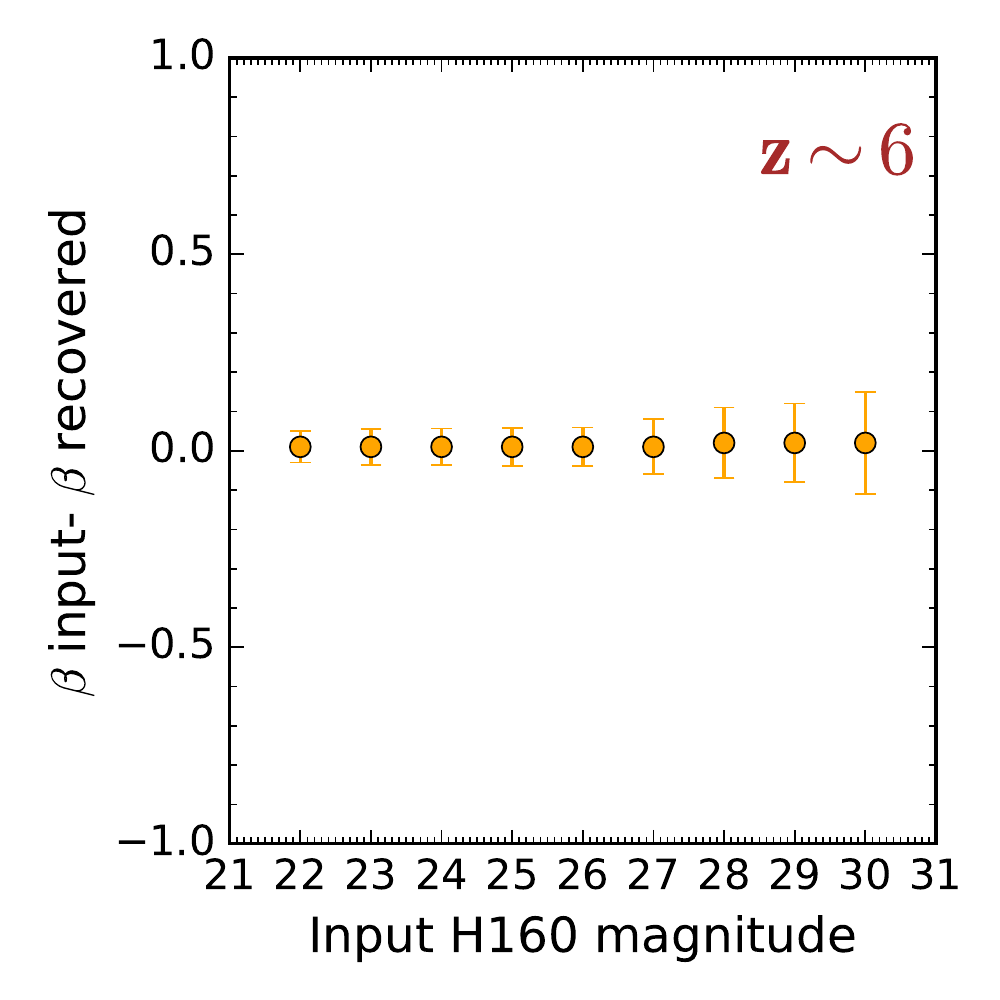}
\end{minipage}
\begin{minipage}{0.45\textwidth}
\centering
\includegraphics[width=1\textwidth, height=0.3\textheight]{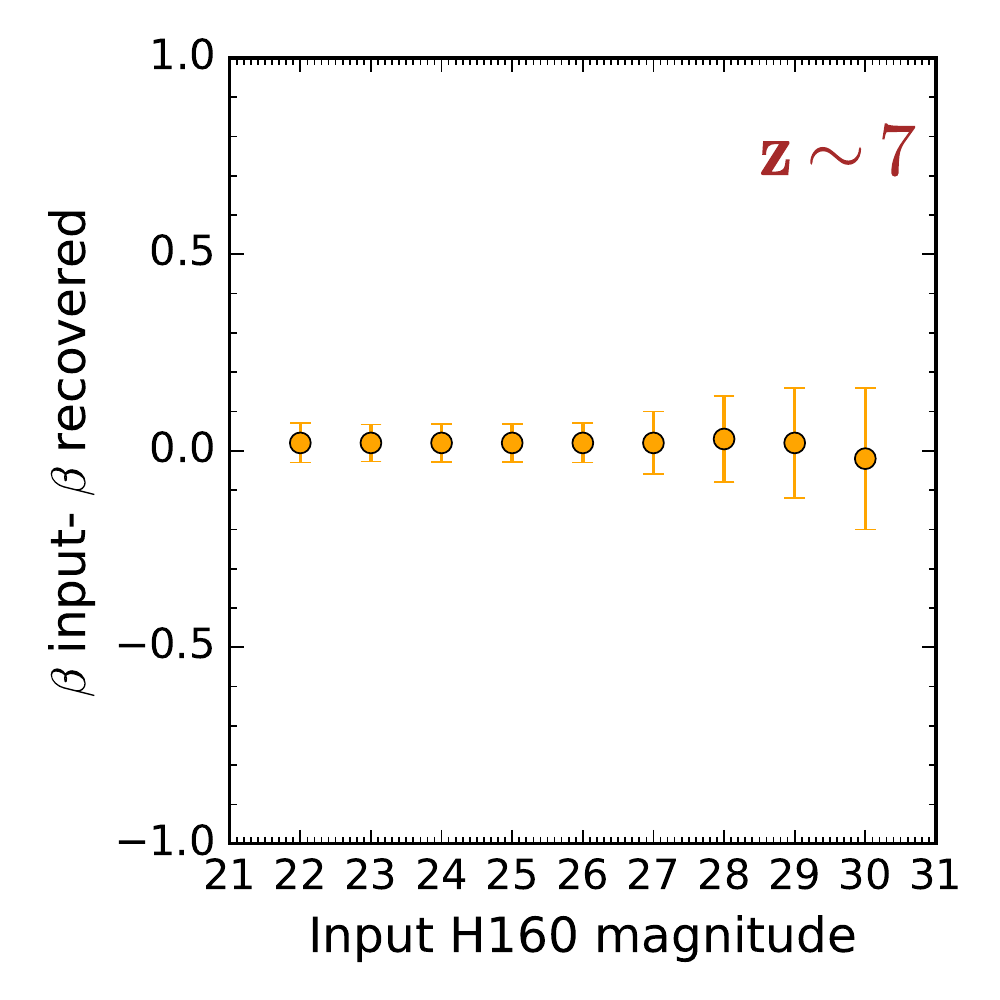}
\end{minipage}
\begin{minipage}{0.45\textwidth}
\centering
\includegraphics[width=1\textwidth, height=0.3\textheight]{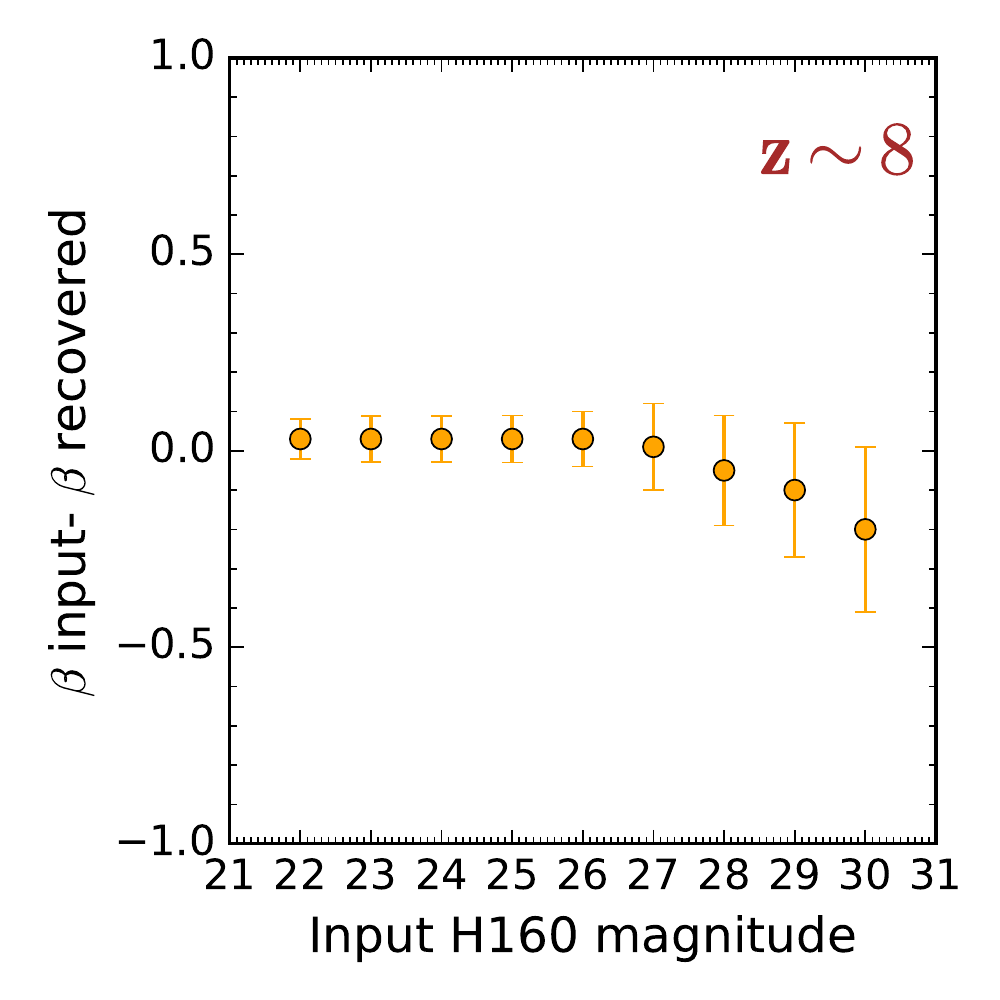}
\end{minipage}
\begin{minipage}{0.45\textwidth}
\centering
\includegraphics[width=1\textwidth, height=0.3\textheight]{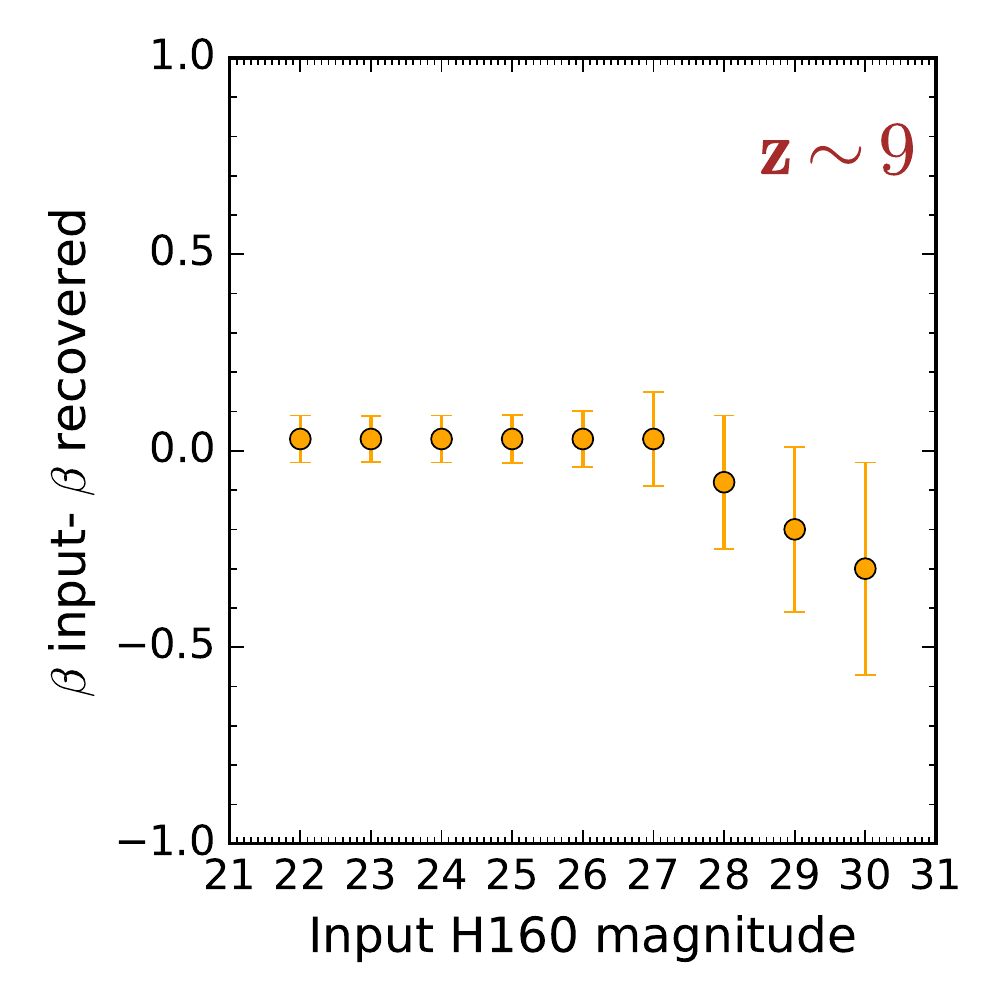}
\end{minipage}
\caption{Results of our simulations at $z=6-9$, showing the difference between the input value of the UV spectral slope $\beta$ and that recovered from SED fitting as a function of input $H_{160}$ magnitude. The orange circles represent the mean difference between input and recovered values of $\beta$ as well as the scatter in bins of $\Delta m=1$. }
\label{fig:betasimulation}
\end{figure*}

Lastly, we run the high redshift sources from the simulated catalog through the same sample selection criteria (See Section~\ref{sec:properties}) as our original sample. We then carry out SED fitting on the galaxies that pass the selection criteria to measure $\beta$ as we did on our real sample. Fig.~\ref{fig:betasimulation} shows the results of these simulations at $z=6-9$, showing the difference between the input value of the UV spectral slope $\beta$ and that recovered the from SED fitting as a function of input $H_{160}$ magnitude. The orange circles represent the mean difference between input and recovered values of $\beta$ as well as the scatter in bins of $\Delta m=1$. As expected, through these simulations we find that the scatter increases at higher redshifts and at faint magnitudes (See Fig.~\ref{fig:betasimulation}), but is generally quite good at recovering the correct values of $\beta$.

\section{Results}\label{sec:results}
\subsection{Correlation between rest frame UV magnitude and UV colours $\beta$ }\label{sec:uvmag_betacorr}
As seen from Fig.~\ref{fig:uv_beta}, previous studies investigating the rest-frame UV colours of galaxies at high redshifts were limited to faint sources at $M_{\mathrm{UV}}<-17$ (shown by the dashed blue line in Fig.~\ref{fig:uv_beta} in $z=6$ bin). Our sample allows us, for the first time, to probe a wide range of magnitudes at $-22<\mathrm{M_{UV}<-13.5}$ at $z=6$. While there are large uncertainties on sources fainter than $M_{\mathrm{UV}}>-17$ due to lensing (shown by a representative error bar at the bottom left of each panel in Fig.~\ref{fig:uv_beta}), it is clear that we do not find a correlation between rest-frame UV magnitude $M_{1500}$ and $\beta$ at $z=6$ (Kendall $\tau=-0.33$). We note that at $z=7$, $z=8$ and $z=9$ the dynamic range in $M_{\mathrm{UV}}$ is reduced. Although, the fact that that we are finding galaxies as faint as $M_{\mathrm{UV}}=-13.5$ at $z=6$ strongly implies that we will be able to these faint UV galaxies at $z>6$ with JWST. Nevertheless, from the current data there appears to be no correlation between rest-frame UV magnitude $M_{1500}$ and $\beta$ at $z=7$ and $z=9$ but there appears to be some correlation at $z=8$, such that fainter galaxies have bluer values of $\beta$.

To quantify this, we first check whether our data has a Gaussian distribution. There are a range of normality tests like the  Shapiro-Wilk or Anderson-Darling test of the residuals that can tell whether the data is unlikely to have come from a normal distribution. However, if the test if not significant, that does not necessarily mean that the data came from a normal distribution or vice-versa. It could also mean that we just do not have enough power to see the difference. Larger sample sizes give more power to detect the non-normality. For small sample sizes such as ours, quantile-quantile (Q-Q) plots can be good diagnostics. Thus, we use Q-Q plots to check for skewness and as there was not too much observed skewness in our data, we choose to fit a first-order polynomial through the median data points at each redshift to estimate the slope and its associated uncertainty. The values of the slope along with its uncertainty are listed in Table~\ref{tab:uv_beta_table}. Although there may appear some correlation at $z=8$, from Table~\ref{tab:uv_beta_table} we conclude that there is no significant correlation ($<2\sigma$) between rest-frame UV magnitude $M_{1500}$ and $\beta$ at all redshifts probed in this study. We also test whether the choice of IMFs has any effect on the estimated rest-frame UV magnitudes and thus on the relation between $\beta$ and $M_{\mathrm{UV}}$. We test this at $z=6$ where the measurement is easiest and find that the rest-frame UV magnitudes get slightly fainter when using a Salpeter IMF. This is consistent with \citet{Jerabkova2017} in which they compare different fluxes with variable IMF. Our results are understandable as Salpeter would produce a relative over-abundance of low-mass stars and consequently an under-abundance of high-mass stars in comparison to Kroupa/Chabrier if the systems have the same total stellar mass. However, we find that difference between rest-frame UV magnitudes with Chabrier and Salpeter is very small ($\leq0.15$) and therefore does not affect the $\beta$-$M_{\mathrm{UV}}$ relation in this work. The difference in magnitudes likely will be more pronounced when the most massive stars are still alive (i.e., in the first few Myr).

At $z=6$, our results are in agreement with \citet{Dunlop2012} and \citet{Finkelstein2012} but in disagreement with \citet{Wilkins2011} who use near-infrared (near-IR) imaging to measure the rest-frame UV continuum colours of galaxies at $4.7<z<7.7$ using a single-colour technique and find lower luminosity galaxies to be bluer than higher luminosity galaxies. Our results are also in disagreement with \citet{Bouwens2012} who determined the UV continuum slopes at $z\sim4-7$ by fitting a power law to the observed photometry and report a well-defined rest-frame UV colour--magnitude relationship that becomes systematically bluer towards fainter UV luminosities. Similarly, more recently, \citet{Bouwens2014b} found a significant colour magnitude relation, such that fainter galaxies displayed bluer slopes, with the relation steepening at $z=4-8$ and our results are in disagreement with them. 

In a similar way, at $z=7$ our results are in agreement with \citet{Dunlop2012}, \citet{Dunlop2013} and \citet{Finkelstein2012} who do not find a significant correlation between rest-frame UV and UV colours but in disagreement with \citet{Wilkins2011}, \citet{Bouwens2012} and \citet{Bouwens2014b} who report a significant colour magnitude relation. Finally, at $z=8$ our results are in agreement with \citet{Finkelstein2012} and \citet{Dunlop2013} but in disagreement with \citet{Bouwens2014b} who found that fainter galaxies have bluer slopes.

As stated earlier, we follow the method of \citet{Finkelstein2012} and derive our UV slopes using an SED fitting method and also measure $M_{\mathrm{UV}}$ by fitting a 100 \AA -wide top-hat filter centered on 1500 Angstrom (See Section~\ref{sec:betacal}). As pointed out in \citet{Finkelstein2012}, the way we measure $M_{\mathrm{UV}}$ may have an effect on whether any correlation between $\beta$ and $M_{\mathrm{UV}}$ is observed. For example, \citet{Bouwens2012} probe different parts of the rest-frame UV depending on the redshift and observed a correlation between $\beta$ and $M_{\mathrm{UV}}$.  Following the method of \citet{Bouwens2012}, \citet{Finkelstein2012} were able to recover a correlation between $M_{\mathrm{UV}}$ and $\beta$, albeit not as strong as \citet{Bouwens2012}. Similarly, \citet{Finkelstein2012} also found a colour-magnitude relation when they used a single colour to derive $\beta$ and therefore argued that these factors may effect whether a trend may or may not be observed. Thus, these factors could likely be the reason for the observed disagreement between our results and other studies.

\begin{figure}
\includegraphics[width=\columnwidth]{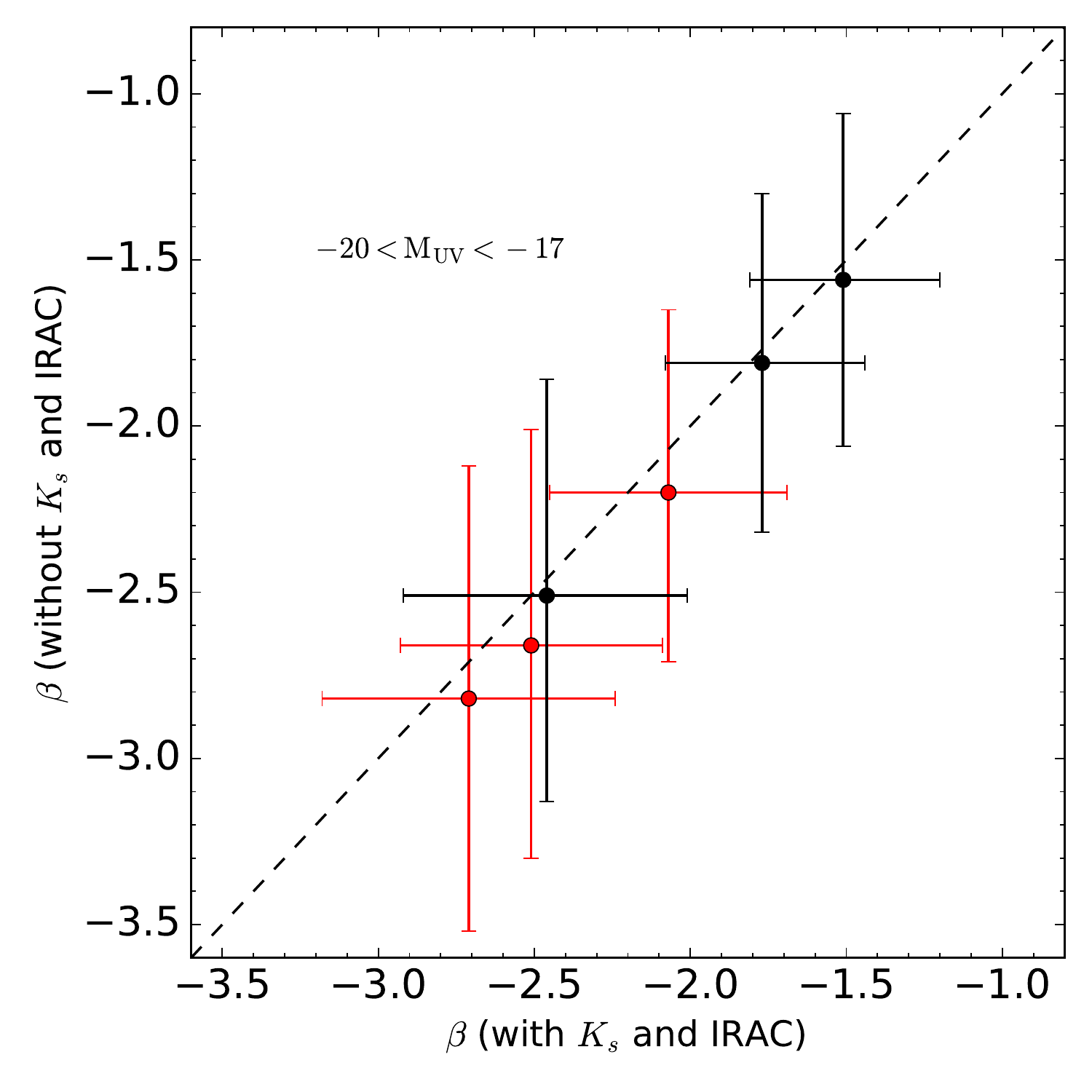}
\caption{Comparison of the two estimated values of $\beta$ with our SED fitting method (with and without longer wavelength data) for a sample of galaxies at $-20<\mathrm{M_{UV}<-17}$. The red circles represent the sources with detections in the longer wavelength channels, whereas the black circles represent the sources with no detections at longer wavelengths.}
\label{fig:beta_comp}
\end{figure}

However, while in this study we choose to use the SED fitting method to derive $\beta$, we test whether deriving $\beta$ with and without the longer wavelength data has any effect on the estimated values of $\beta$. This test is motivated by the fact that while \citet{Finkelstein2012} have shown with the help of simulations that SED fitting is a superior choice over power-law and single-colour method, they have reported error bars in excess of unity on $\beta$ as they were affected by the lack of VLT or Spitzer data. Therefore, since we use all the photometric bands including VLT and Spitzer in our SED fitting in this work, it is worth checking if these bands indeed have any effect on the estimated values of $\beta$. We explore two scenarios in this case:
\begin{itemize}
\item
Deriving $\beta$ when sources are detected in VLT or Spitzer: In this case we find that if there are detections in longer wavelenghts, and if this data is being used in the SED fitting to derive $\beta$, it invariably leads to better estimates of colors, such that the $\beta$ values are redder with smaller error bars. We find that the error bars are indeed $\geq$ unity when we do not include the longer wavelength data. However, the error bars are reduced by as much as 50$\%$ when longer wavelength information is included. This is shown by red circles in Fig.~\ref{fig:beta_comp}, clearly demonstrating that the estimated values of $\beta$ for sources that are detected in longer wavelengths are redder with smaller uncertainties when the colours are estimated with the inclusion of longer wavelength data.
\item
Deriving $\beta$ when sources are not detected in VLT or Spitzer: It can be argued that using the longer wavelength data would be useful only if there is actually some useful information present in the data (i.e., if sources are detected in them) to constrain the best-fitting SED. We therefore test this for our sample of galaxies that are not detected in longer wavelengths, but by still including this information in our SED fitting. We find that in this case the estimated values of $\beta$ are also redder (albeit not as red had they been detected in VLT or Spitzer) but their error bars are reduced by as much as 35$\%$. This is shown by the black circles in Fig. 4, showing that the estimated $\beta$ values are slightly redder, with smaller uncertainties, when colours are measured with longer wavelengths.
\end{itemize}

This shows that whilst it is important to have real detections in longer wavelengths to notice significant improvement in estimated values of $\beta$ and their uncertainties, we can still derive better estimates of $\beta$ values with reduced uncertainties even if sources are not detected in longer wavelengths but the information is still being used in SED fitting.

We also caution that care has to be taken when there are no detections in the longer wavelengths but this information is still being used. We find that in such cases this can lead to biased bluer values of $\beta$, if the photometric measurements are unrealiable. For example, as identified in \citet{Bhatawdekar2019}, when doing photometry with T-PHOT on clusters, if bright sources are near the faint sources, then the photometry of nearby faint sources is affected by the residuals of the bright sources. In such cases, the recovered S/N from T-PHOT is significantly high or significantly negative even if the sources are not present in the longer wavelength channels, suggesting that their photometry is unreliable. We find that if photometric measurements of such contaminated sources are used,  they not only cause some unfortunate high-z solutions (as noted in \citet{Bhatawdekar2019}) but the derived $\beta$ values of such sources will also be biased blue. We therefore carefully inspect the photometric measurements of such sources and our final sample consists of only those sources whose photometric measurements are reliable, leading to more robust estimates of $\beta$.

\subsection{Correlation between $\beta$ and redshift}
\begin{figure}
\includegraphics[width=\columnwidth]{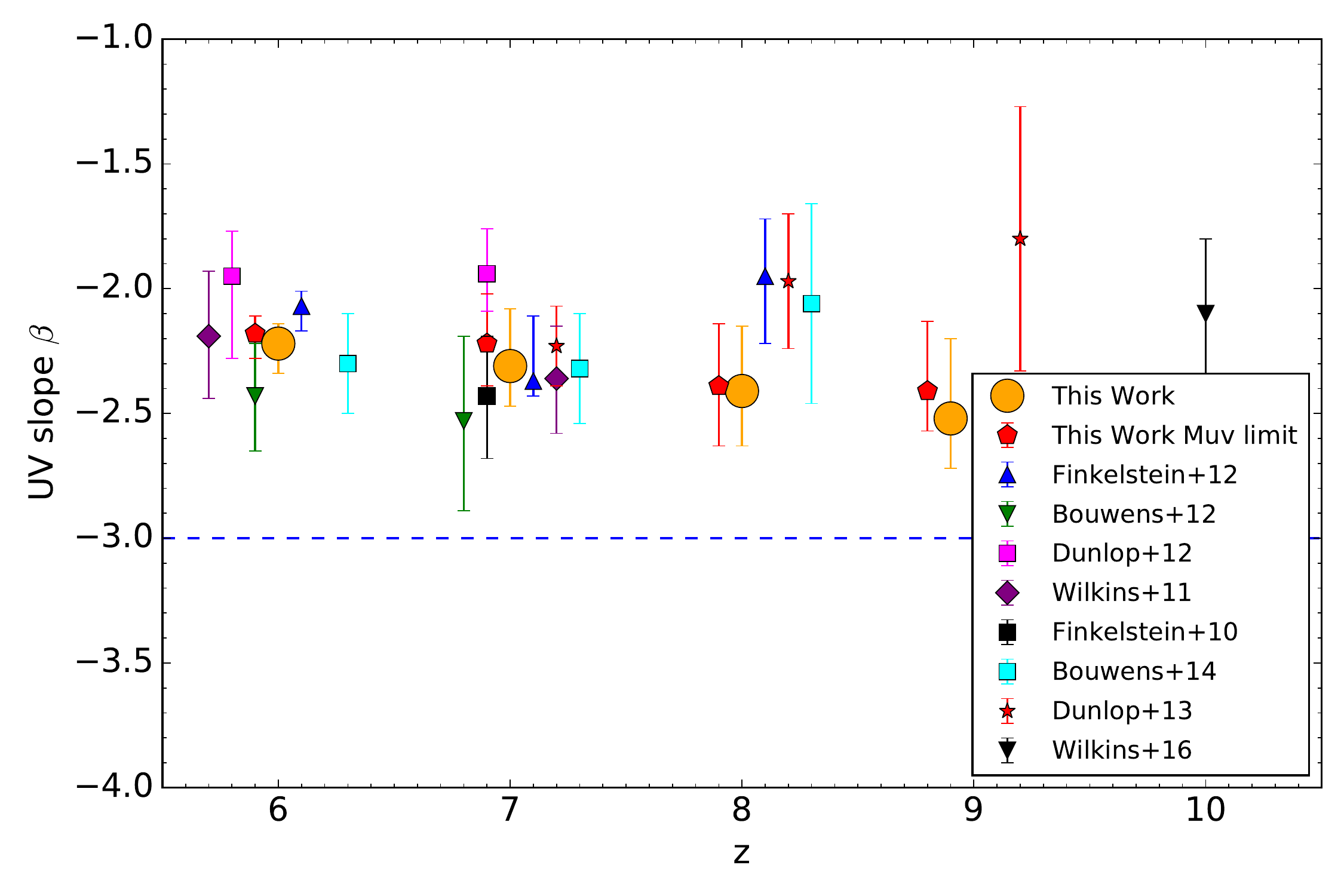}
\caption{The evolution of the median $\beta$ of all galaxies with redshift shown in yellow circles and the evolution of mean $\beta$ of all galaxies at a magnitude cut of $M_{\mathrm{UV}}=-18$, in red pentagons. The median $M_{\mathrm{1500}}$ for these points are $-17.96$, $-18.53$, $-18.77$, and $-19.44$ at $z=6$, $z=7$, $z=8$, and $z=9$ respectively. Shown also are the results from the literature interpolated at the median value of $M_{1500}$ for our sample, except the $z = 9$ result of \citet{Dunlop2013} shown at $M_{1500}=-18$.  The blue dashed line suggests the expected colours if galaxies had stars with very-low metallicities ($\sim10^{-2}\mathrm{Z_{\odot}}$.)}
\label{fig:muv_beta_evolution}
\end{figure}

From Fig.~\ref{fig:uv_beta}, we find that there is no significant correlation of $\beta$ with UV luminosity. We now therefore examine if there is any evolution of $\beta$ with redshift by plotting the median $\beta$ values of all galaxies as a function of redshift. To do this, we first interpolate the results from previous literature at the median value of $M_{1500}$ of our sample except the $z = 9$ result of \citet{Dunlop2013} shown at $M_{1500}=-18$. As seen from Fig.~\ref{fig:muv_beta_evolution}, $\beta$ appears to evolve mildly from $-2.22_{-0.12}^{+0.08}$ at $z\sim6$ to $-2.52_{-0.20}^{+0.32}$ at $z\sim9$, presumably due to rising dust extinction. To check if faint galaxies in our sample are driving this mild evolution, we also compute the average value of $\beta$ at each redshift at a upper magnitude cut of $M_{\mathrm{UV}}=-18$, shown in red pentagons in Fig.~\ref{fig:muv_beta_evolution}. As seen from Fig.~\ref{fig:muv_beta_evolution}, the $\beta$ values are redder when using this magnitude limit and we find no significant slope ($<2\sigma$) between $\beta$ and redshift, suggesting that faint galaxies in our sample are likely driving the apparent evolution. 

Comparing our estimated values of $\beta$ with previous work, such as \citealt{Finkelstein2012, Bouwens2012, Bouwens2014b}, we find that our results are in agreement with them, such that galaxies on average get bluer at higher redshift. However, our results are not in agreement with \citet{Dunlop2012} who studied the UV colours of galaxies at $z>6$ in the HUDF, ERS and HUDF09-2 fields, and reported an average value of $\beta=-2.05\pm0.1$ at all redshifts. Our $\beta$ values are bluer than theirs at all redshifts. We note that they apply a stringent selection criteria and restrict their sample to contain objects that have at least one $8\sigma$ near-IR detection in the WFC3/IR data, which may have excluded many of the faint galaxies, causing $\beta$ to have a redder value. We therefore test this by applying the same cut on our sample and find that our $\beta$ changes from $\beta=-2.20_{-0.10}^{+0.10}$ at $z\sim6$ to $\beta=-2.51_{-0.21}^{+0.30}$ at $z\sim9$ when we restrict our sample with the same criteria as \citet{Dunlop2012}. With these results we conclude that applying a stringent criteria does not cause $\beta$ to have a redder value. Similarly, our results are also not in agreement with \citet{Dunlop2013} who use the imaging from UDF12 campaign to calculate the UV colours of galaxies at redshifts $z>6.5$ and report an average value of $\beta=-2.1\pm0.2$, $\beta=-1.9\pm0.3$ and $\beta=-1.8\pm0.6$ at $z\sim7$, $z\sim8$ and $z\sim9$ respectively.

At $z\sim6$ our $\beta$ values are bluer than \citet{Wilkins2011}, \citet{Finkelstein2012} and \citet{Dunlop2012} but redder than \citet{Bouwens2012}. Comparing at $z\sim7$, our estimated UV slopes are redder than \citet{Finkelstein2010}, \citet{Finkelstein2012}, \citet{Bouwens2012} and \citet{Wilkins2011} but bluer than \citet{Dunlop2012}, \citet{Dunlop2013} and \citet{Bouwens2014b}. At $z\sim8$ our $\beta$ values are bluer than \citet{Finkelstein2012} and \citet{Dunlop2013} and \citet{Bouwens2014}, and finally at $z\sim9$ our UV slopes are bluer than \citet{Dunlop2013}. As stated in Section~\ref{sec:uvmag_betacorr}, these differing results are likely due to our use of all the photometric bands, including data from VLT and \textit{Spitzer}, yielding more robust estimates of $\beta$. In Fig.~\ref{fig:muv_beta_evolution}, the blue dashed line suggests the expected colours if galaxies had stars with very-low metallicities ($\sim10^{-2}\mathrm{Z_{\odot}}$). Although our estimated value of $\beta$ has large uncertainties at $z\sim9$, our results show that the UV colours of galaxies at the highest redshifts probed with HFF are not blue enough to have stars with very low metallicities and only JWST will be able to provide a clear picture of this by probing higher redshifts and deeper magnitudes.

\subsection{$\beta$ of faintest galaxies}
In this section we compare the $\beta$ values for the faintest galaxies as these are the systems that are thought to have unusual spectra or are responsible for reionizing the Universe. At $z\sim7$, \citet{Bouwens2010} measured $\beta$ for their sample of galaxies, finding that the very low luminosity galaxies exhibited UV continuum slopes as steep as $\beta=-3$ and argued the likelihood of the presence of extremely metal-poor stars or a top--heavy IMF in these galaxies, supporting that such exotic populations might be crucial to yield such blue values. \citet{Finkelstein2010} also reported similar steep values of $\beta$ at $z\sim6-7$ by examining the same dataset, albeit with larger uncertainties, and therefore concluded that exotic populations were not necessary for such bluer values of the UV slope $\beta$. Furthermore, with the help of simulations, \citet{Dunlop2012} showed that there is a bias towards artificially blue slopes for faint galaxies and argued that the very blue colours are likely overestimated. To look into this in more detail, with the help of an improved dataset and with the SED fitting method, \citet{Finkelstein2012} re-examined this and reported a value of $\beta=-2.68_{-0.24}^{+0.39}$ for faint galaxies at $z\sim7$, redder than their previously reported results. Similarly, more recently \citet{Bouwens2014b} measured the UV continuum slopes of their sample of galaxies at $z=4-8$ and reported that their $\beta$ values are redder than their previously reported values ($\beta=-2.42\pm0.28$ at $z=7$).

In this work, we employ the SED fitting method similar to \citet{Finkelstein2012} but also use all the photometric bands, including data from VLT and \textit{Spitzer}, to derive more robust values of UV slopes for our sample of galaxies. As shown in Table~\ref{tab:uv_beta_table}, at $z\sim7$ the bluest value of our sample is $\beta=-2.32_{-0.23}^{+0.30}$, which is redder than \citet{Finkelstein2012}, \citet{Bouwens2010} and \citet{Bouwens2014}. Similarly, at $z\sim9$ we find that our bluest data point has a value of $\beta=-2.63_{-0.43}^{+0.52}$ at $z\sim9$ (See Table~\ref{tab:uv_beta_table}), finding no evidence as of yet for unusual or Pop III stellar populations (with values $\beta\leq-3$) at $z>6$ with HFF.

\subsection{Correlation of $\beta$ with stellar mass}\label{sec:betamass}
\begin{figure*}
\centering
\begin{minipage}{0.45\textwidth}
\centering
\includegraphics[width=1\textwidth, height=0.3\textheight]{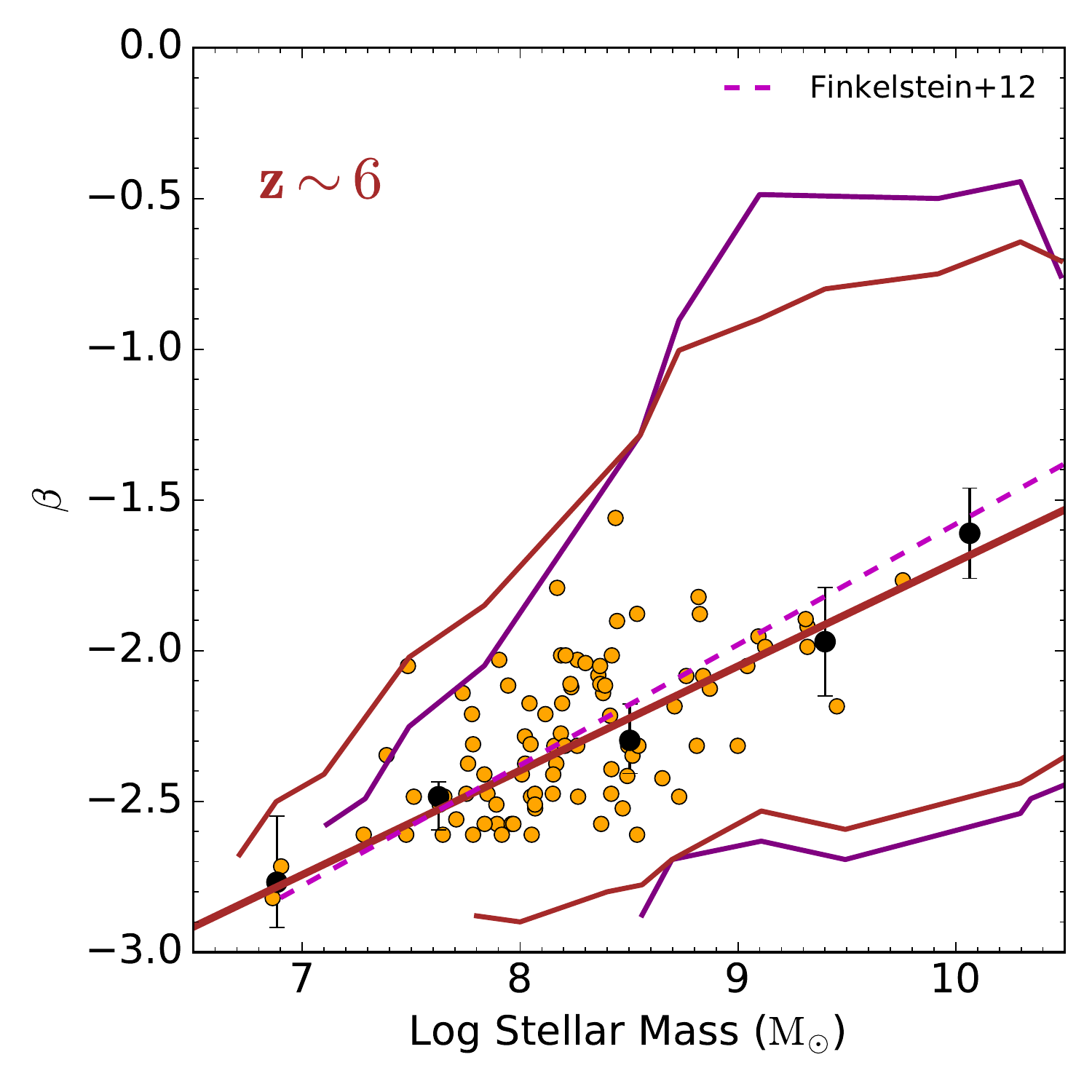}
\end{minipage}
\begin{minipage}{0.45\textwidth}
\centering
\includegraphics[width=1\textwidth, height=0.3\textheight]{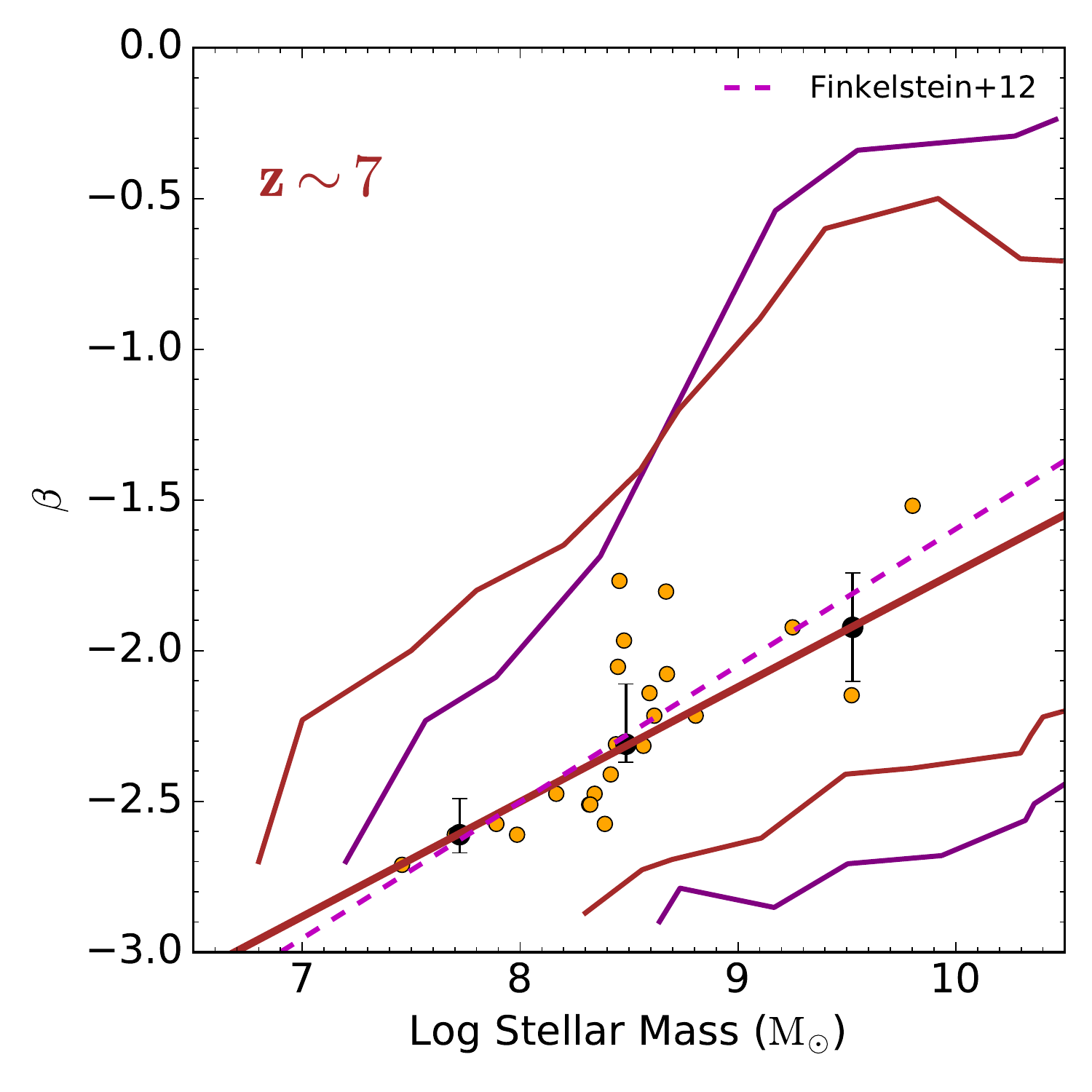}
\end{minipage}
\begin{minipage}{0.45\textwidth}
\centering
\includegraphics[width=1\textwidth, height=0.3\textheight]{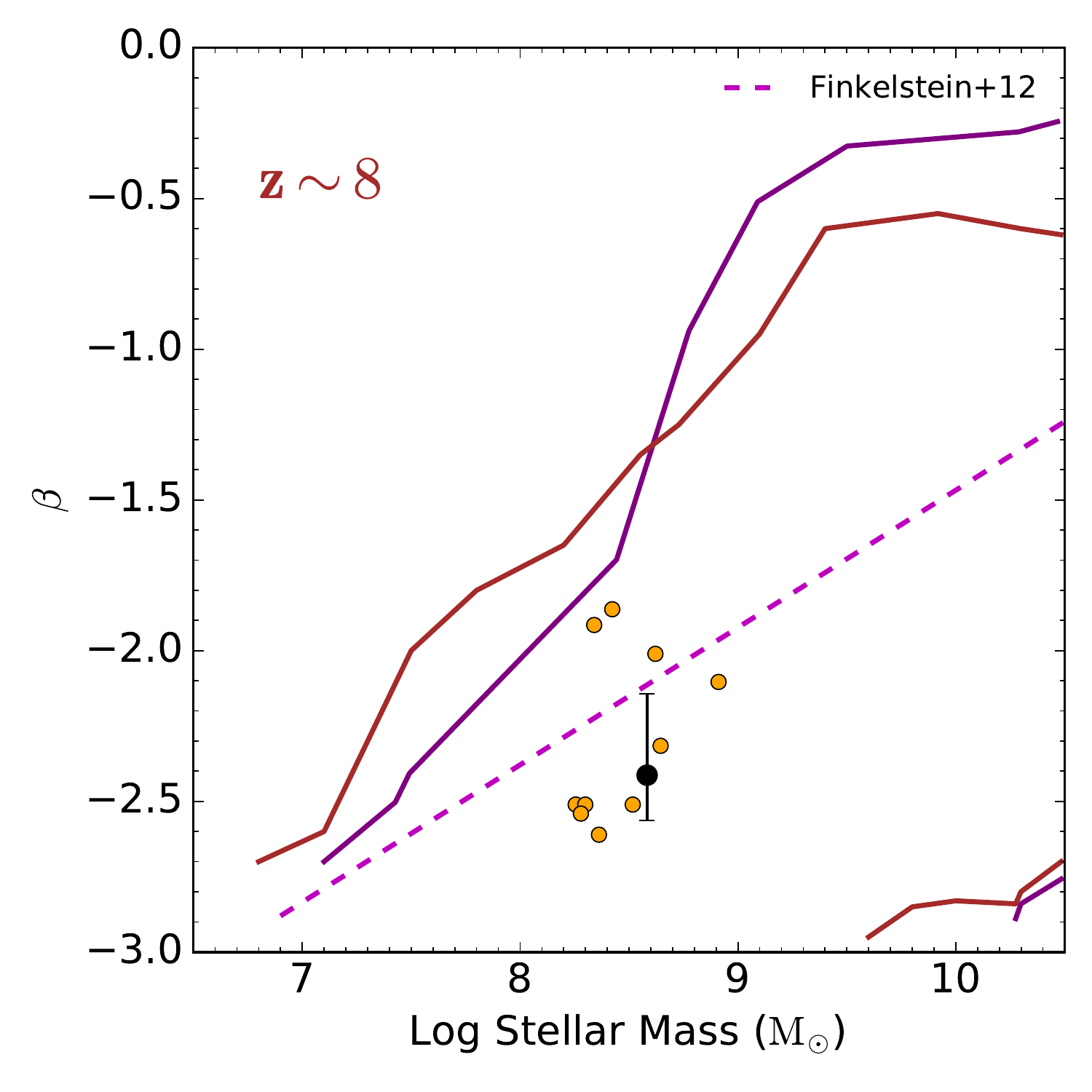}
\end{minipage}
\begin{minipage}{0.45\textwidth}
\centering
\includegraphics[width=1\textwidth, height=0.3\textheight]{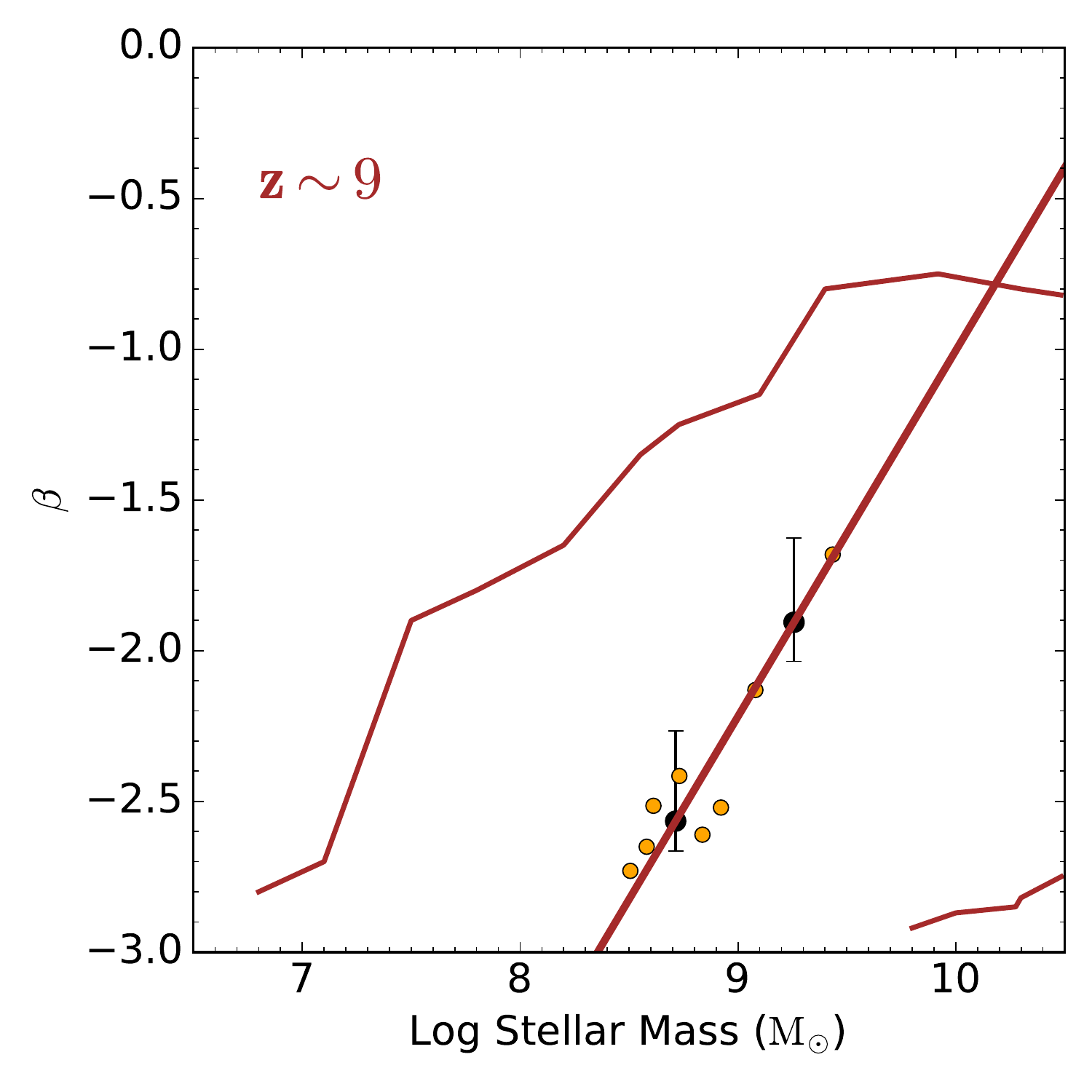}
\end{minipage}
\caption{Estimated UV slope $\beta$ vs stellar mass at $z=6-9$. The filled yellow circles show the results for individual galaxies, whereas the black circles show the median value of $\beta$ in each stellar mass bin of $1M_{\odot}$, with the uncertainties being the errors on the median, estimated with bootstrap Monte Carlo simulations. The solid red lines shows a linear fit through the median $\beta$ points. There is a strong correlation between $\beta$ and stellar mass in all redshift bins, such that lower mass galaxies exhibit bluer UV slopes. The best-fit lines of \citet{Finkelstein2012} are also shown by dashed magenta lines for comparison. The red curve denotes the 20 per cent completeness level estimated from our simulations, whereas the purple curve is the 20 per cent completeness level from \citet{Finkelstein2012} shown for comparison.}
\label{fig:mass_beta}
\end{figure*}

In Section~\ref{sec:uvmag_betacorr} we determined that there is no strong correlation of $\beta$ with UV luminosity. In this section, we therefore proceed to determine whether there exists any trend between $\beta$ and stellar mass. To do this, we use the wide dynamic range in stellar masses ($10^{6.8}-10^{10}M_{\odot}$) estimated for our sample of high redshift galaxies at $z=6-9$ from \citet{Bhatawdekar2019}, and plot $\beta$ as a function of redshift, as shown in Fig.~\ref{fig:mass_beta}. Here, we compute the median values of $\beta$ in different mass bins of $6<\mathrm{log}M/M_{\odot}<7$, $7<\mathrm{log}M/M_{\odot}<8$, $8<\mathrm{log}M/M_{\odot}<9$, $9<\mathrm{log}M/M_{\odot}<10$ and $10<\mathrm{log}M/M_{\odot}<11$, similar to \citet{Finkelstein2012}. The median values of $\beta$ are as listed in Table~\ref{tab:mass_beta_table} along with the associated uncertainties, which were estimated by bootstrap simulations described in Section~\ref{sec:betacal}.

 \begin{deluxetable*}{cccccccc}
\tabletypesize{\small}
\tablecaption{Median values of the UV spectral slope $\beta$ in different mass bins}
\label{tab:mass_beta_table}
\tablewidth{0pt}
\tablehead{
\colhead{$z$} & \colhead{Median $\beta$} & \colhead{Median $\beta$} & \colhead{Median $\beta$} & \colhead{Median $\beta$} & \colhead{Median $\beta$} & \colhead{$\beta$--Stellar Mass} \\[-0.1cm]
\colhead{$ $} & \colhead{$\log M/M_{\odot}=$} & \colhead{$\log M/M_{\odot}=$} & \colhead{$\log M/M_{\odot}=$} & \colhead{$\log M/M_{\odot}=$} & \colhead{$\log M/M_{\odot}=$} & \colhead{Slope}\\[-0.1cm]
\colhead{$ $} & \colhead{$6-7$} & \colhead{$7-8$} & \colhead{$8-9$} & \colhead{$9-10$} & \colhead{$10-11$} & \colhead{}
}
\startdata
 6  & $-2.77_{-0.15}^{+0.22}$ & $-2.48_{-0.11}^{+0.04}$ & $-2.30_{-0.10}^{+0.07}$ & $-1.97_{-0.13}^{+0.18}$ & $-1.61_{-0.11}^{+0.15}$ & $0.34\pm0.05$\\
		 7 & $..$ & $-2.61_{-0.05}^{+0.12}$ & $-2.31_{-0.11}^{+0.20}$ & $-1.92_{-0.20}^{+0.18}$ & $..$ & $0.38\pm0.06$\\
		 8  & $..$ & $..$ & $-2.41_{-0.15}^{+0.27}$ & $..$ & $..$ & $..$\\
		 9  & $..$ & $..$ & $-2.56_{-0.09}^{+0.30}$ & $-1.90_{-0.13}^{+0.28}$ & $..$ & $1.21\pm0.50$\\
\enddata
\end{deluxetable*}

As seen in Fig.~\ref{fig:mass_beta}, our sample allows us to probe a wide range of stellar masses at $z=6$ and a strong correlation between $\beta$ and stellar mass is apparent at this redshift, such that lower mass galaxies exhibit bluer UV slopes. At $z=7$, $z=8$ and $z=9$ although the dynamic range in stellar mass is reduced, the current data shows a correlation between $\beta$ and stellar mass. 

To quantify this, we fit a first-order polynomial through the median data points at each redshift to estimate the slope and its associated uncertainty. The best-fit line at each redshift, except $z=8$ since there is only one median data point, is shown Fig.~\ref{fig:mass_beta}, and the values of the slope along with its uncertainty are listed in Table~\ref{tab:mass_beta_table}. We also show the best-fit lines of \citet{Finkelstein2012} in Fig.~\ref{fig:mass_beta} for comparison. As seen from Table ~\ref{tab:mass_beta_table}, we find a $>5\sigma$ significance dependence between $\beta$ and stellar mass at $z=6$ and $z=7$ and a $>2\sigma$ significance at $z=9$. We also check whether the choice of IMFs has any effect on the observed correlation between $\beta$ and stellar mass. To do this we use the Salpeter IMF to remeasure our $\beta$ values at $z=6$ and find that, as expected, the stellar masses are higher by 0.24 dex. This is understandable as for a given total mass of a stellar system, the rest-frame magnitudes would get fainter for Salpeter IMF relative to Chabrier/Kroupa. However, if we observe a system with a given UV luminosity, that is with observationally fixed high-mass stellar content dominating the UV emission, the inferred star-formation rate and stellar mass of this system would be higher when using a Salpeter IMF. This is because for the same amount of high mass stars, Salpeter contains a larger number of low-mass stars, which then inflates the total stellar mass and thus also the star-formation rate relative to Chabrier/Kroupa. Therefore, with a Salpeter IMF our $\beta$-stellar mass correlation slope changes slightly from $0.34\pm0.05$ at $z=6$ to $0.32\pm0.05$.

To check if our observed $\beta$ and stellar mass relation is affected by incompleteness, we calculate the ratio of the number of recovered galaxies to the number of input galaxies from our simulations performed in Section~\ref{sec:testingmethod}. The estimated 20 per cent completeness level is shown in Fig.~\ref{fig:mass_beta} with the red curve. For comparison, we also show the completeness curves derived by \citet{Finkelstein2012} in purple. Inspecting the plot, it appears that we would have discovered red low mass galaxies ($-2.0<\beta<-1.5$, and $7.5<\mathrm{log}M/M_{\odot}<8.5$) if they were present and therefore we conclude that our stellar mass-$\beta$ relation is true.

\begin{figure}
\includegraphics[width=\columnwidth]{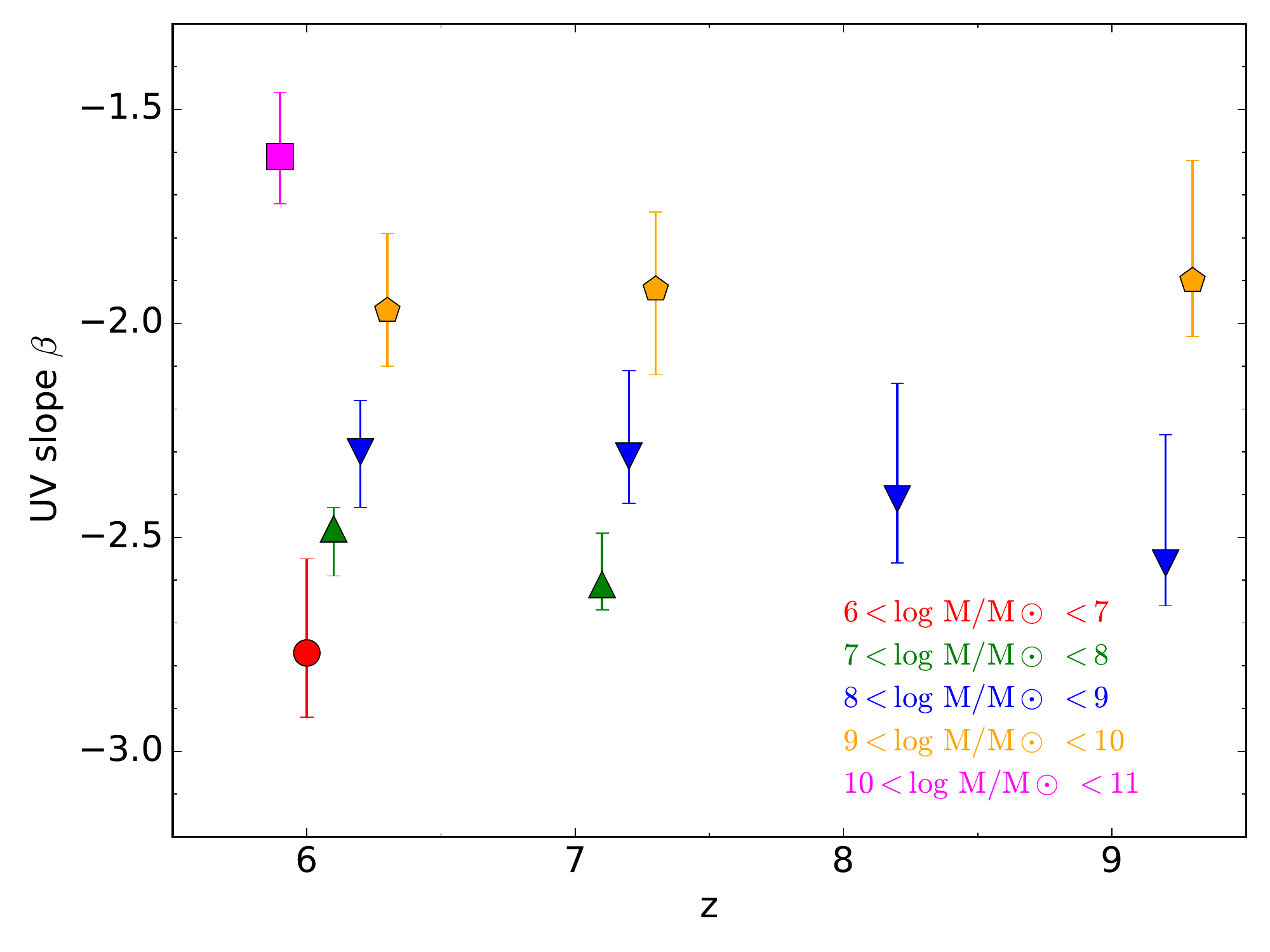}
\caption{Evolution of the median value of UV spectral slope $\beta$ in different mass bins with redshift. The red circle shows the median $\beta$ in mass bin $6<\mathrm{log}M/M_{\odot}<7$, green triangles show the median $\beta$ in mass bin $7<\mathrm{log}M/M_{\odot}<8$, blue triangles show the median $\beta$ in mass bin $8<\mathrm{log}M/M_{\odot}<9$, yellow pentagons show the median $\beta$ in mass bin $9<\mathrm{log}M/M_{\odot}<10$ and finally the magenta square shows the median $\beta$ in mass bin $10<\mathrm{log}M/M_{\odot}<11$. It appears that low mass galaxies at $\log M/M_{\odot}<9$ become bluer with increasing redshift, whereas the massive galaxies at $\log M/M_{\odot}>9$ appear to exhibit a nearly constant $\beta$ at each redshift.  }
\label{fig:mass_beta_evolution}
\end{figure}

To investigate this further, we once again plot $\beta$ as a function of redshift, by spliting our sample into the same mass bins; $6<\mathrm{log}M/M_{\odot}<7$, $7<\mathrm{log}M/M_{\odot}<8$, $8<\mathrm{log}M/M_{\odot}<9$, $9<\mathrm{log}M/M_{\odot}<10$ and $10<\mathrm{log}M/M_{\odot}<11$, as shown in Fig.~\ref{fig:mass_beta_evolution}. Examining the plot, it appears that low mass galaxies at $\log M/M_{\odot}<9$ become bluer with increasing redshift, whereas the massive galaxies at $\log M/M_{\odot}>9$ appear to exhibit approximately constant $\beta$ at each redshift. \citet{Finkelstein2012} notice a similar effect and suggest that this is likely because feedback from supernovae explosions is driving the dust out of low mass galaxies whereas massive galaxies are able to retain this dust due to their higher gravitational potential.

\subsection{Correlation of $\beta$ with SFR}
\label{sec:betasfr}
\begin{figure*}
\centering
\begin{minipage}{0.45\textwidth}
\centering
\includegraphics[width=1\textwidth, height=0.3\textheight]{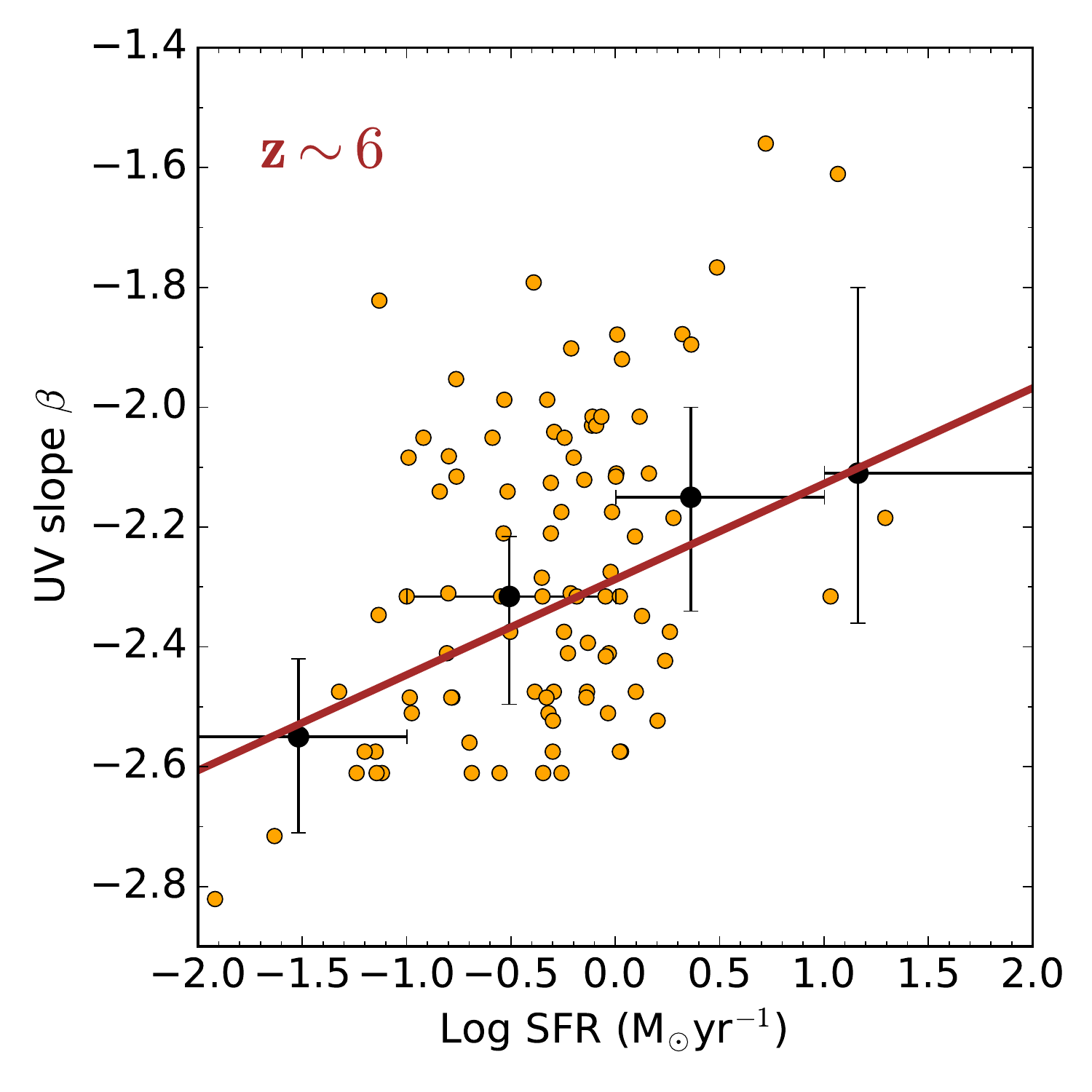}
\end{minipage}
\begin{minipage}{0.45\textwidth}
\centering
\includegraphics[width=1\textwidth, height=0.3\textheight]{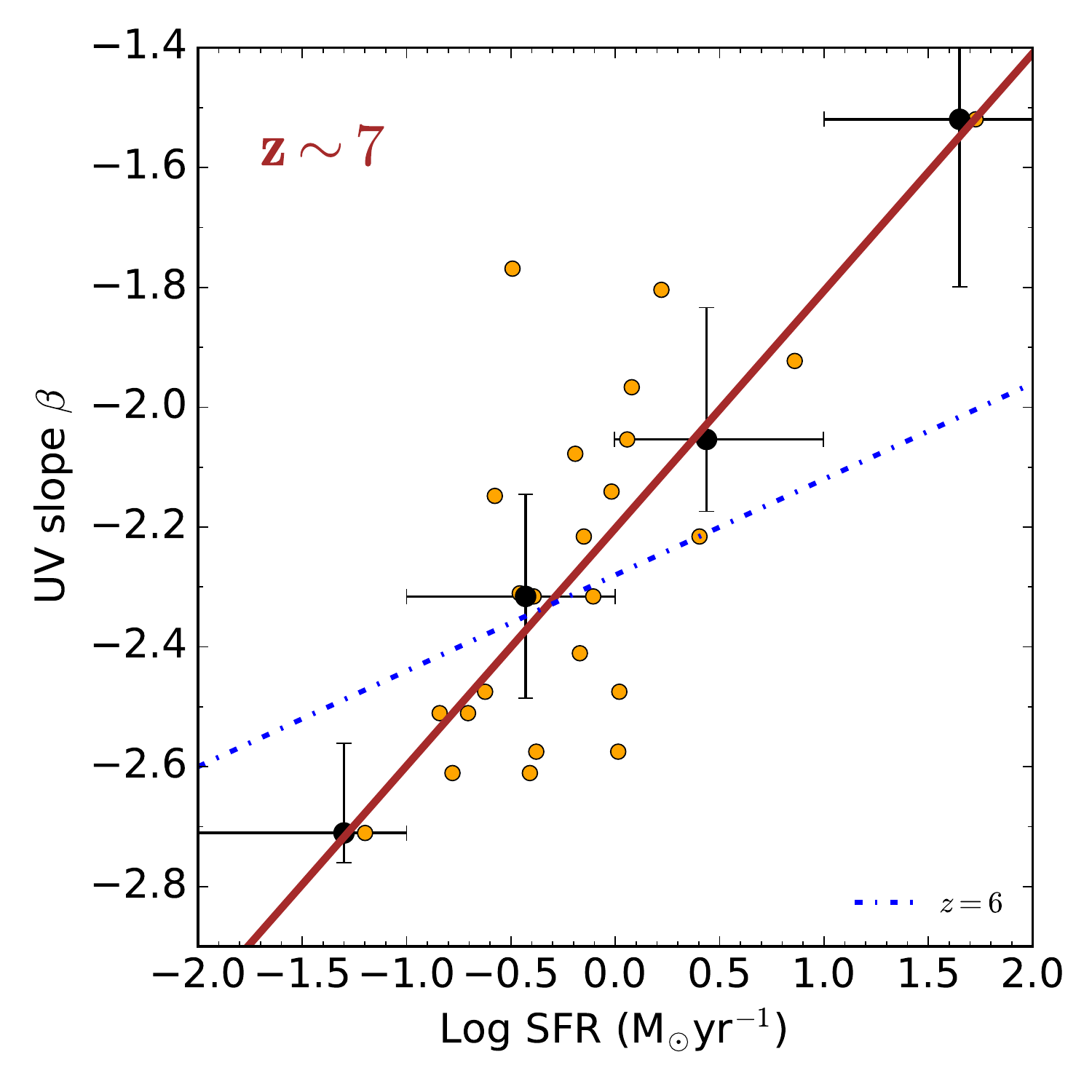}
\end{minipage}
\begin{minipage}{0.45\textwidth}
\centering
\includegraphics[width=1\textwidth, height=0.3\textheight]{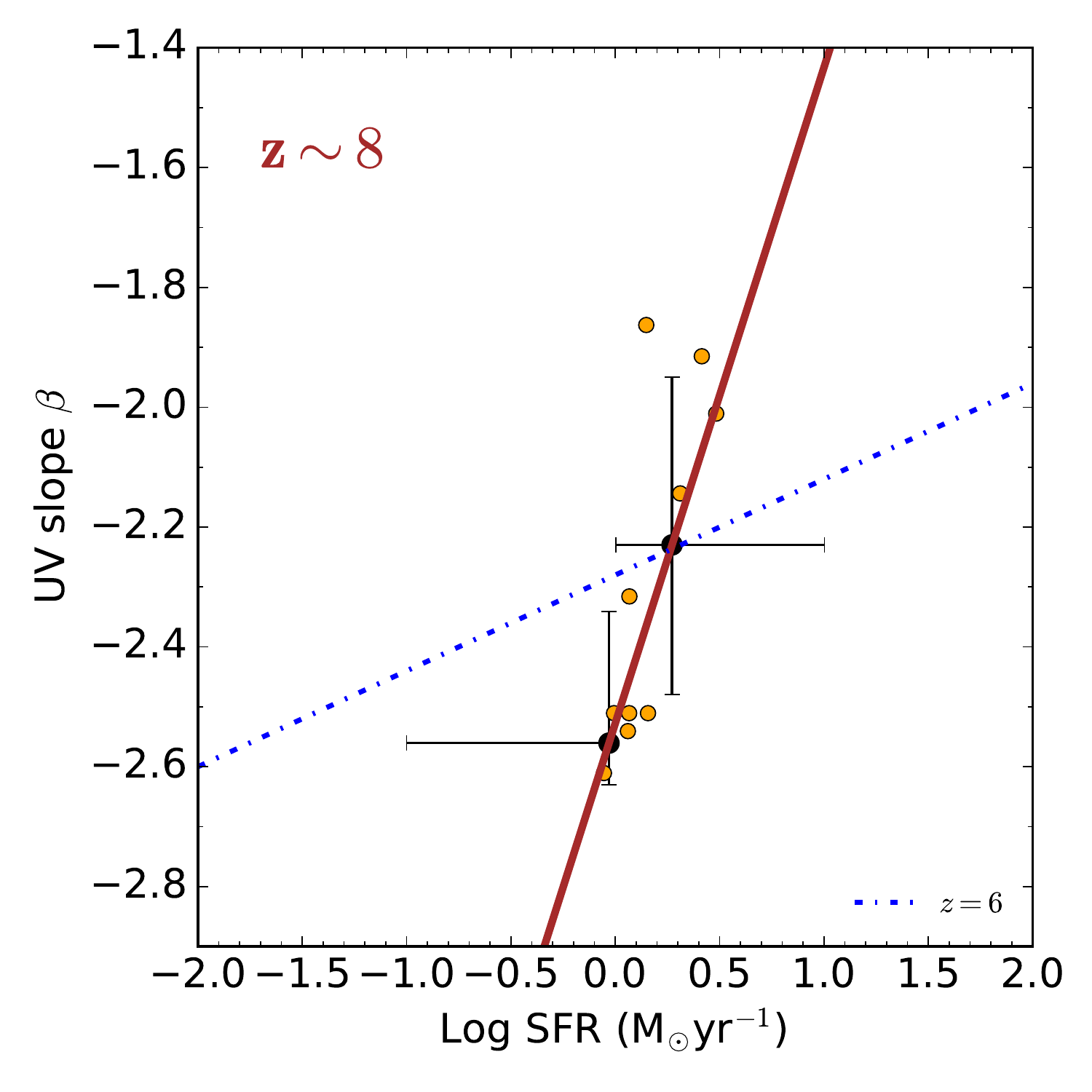}
\end{minipage}
\begin{minipage}{0.45\textwidth}
\centering
\includegraphics[width=1\textwidth, height=0.3\textheight]{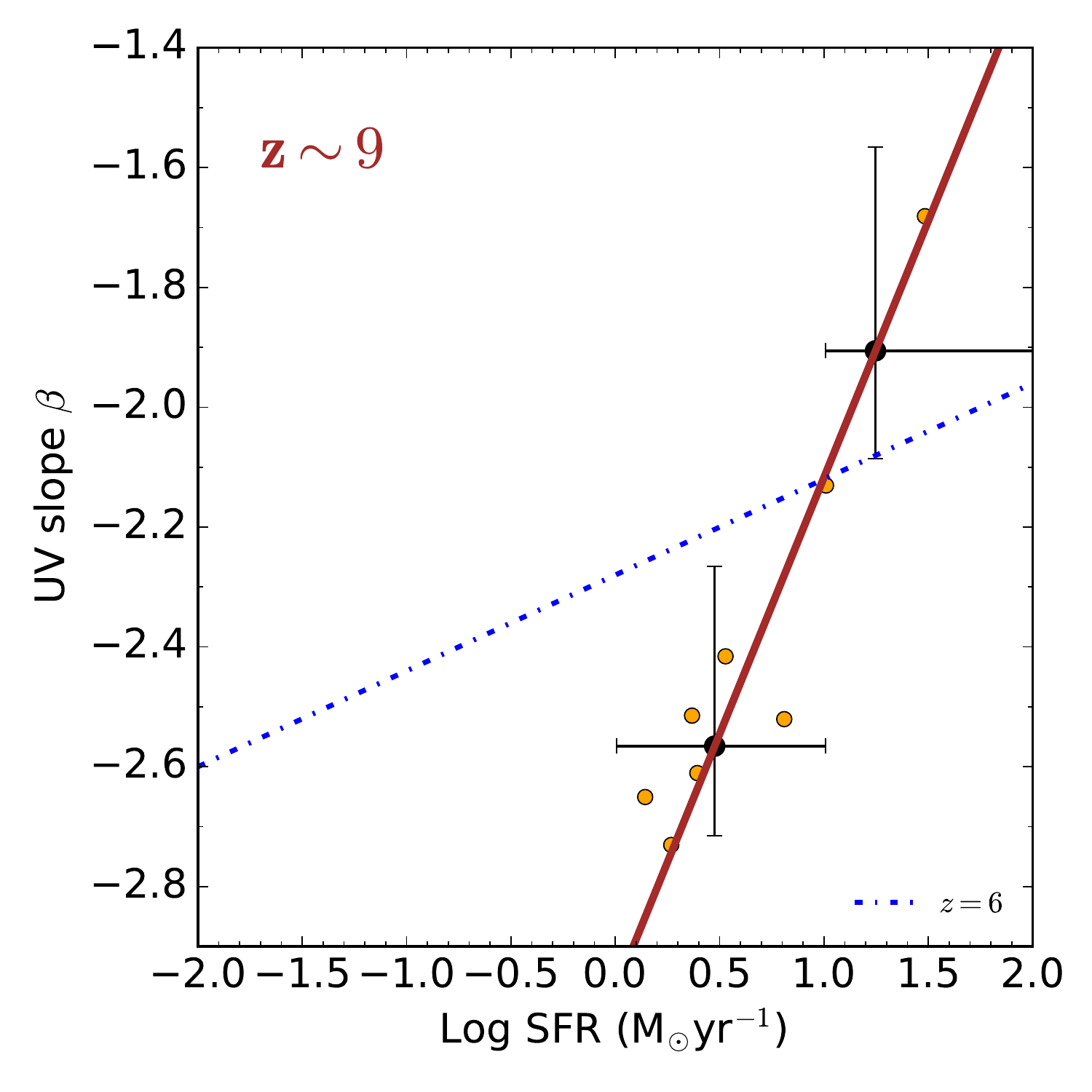}
\end{minipage}
\caption{Measured UV slope $\beta$ vs SFR at $z=6-9$. The filled yellow circles show the results for individual galaxies, whereas the black circles show the median value of $\beta$ in each SFR bin of $1\mathrm{M_{\odot}yr^{-1}}$. The vertical error bars denote the errors on the median estimated with bootstrap Monte Carlo simulations whereas the horizontal error bars represent the width of the bins. The solid red lines show a linear fit through the median $\beta$ points. There is a strong correlation between $\beta$ and SFR, such that galaxies with low SFRs exhibit bluer slopes. The best-fit line at $z=6$ is copied on other redshifts as a reference point, shown by dashed blue line.}
\label{fig:sfr_beta}
\end{figure*}

We now investigate if there exists any correlation between $\beta$ and ongoing star formation by plotting $\beta$ as a function of SFR. To do this, we use the dust corrected star formation rates (further corrected for magnification) from \citet{Bhatawdekar2019}, and compute the median values of $\beta$ of our sample in bins of $-2.0<\log\mathrm{SFR}<-1.0$, $-1.0<\log\mathrm{SFR}<0.0$, $0.0<\log\mathrm{SFR}<1.0$ and  $1.0<\log\mathrm{SFR}<2.0$. This is shown by black circles in Fig.~\ref{fig:sfr_beta}. Similarly, in Table~\ref{tab:sfr_beta_table}, we list the median values in the mentioned SFR bins along with the associated uncertainties estimated with our bootstrap simulations.

Inspecting Fig.~\ref{fig:sfr_beta}, there is a reasonably wide range of SFRs at $z=6$ and $z=7$ and there appears to be a strong correlation between $\beta$ and SFR at these redshifts, such that galaxies with low SFRs exhibit bluer slopes. At $z=8$ and $z=9$ although the dynamic range in SFRs is reduced, a correlation between SFR and $\beta$ is apparent. To quantify this, we fit a first-order polynomial through the median data points at each redshift to estimate the slope and its associated uncertainty. The best-fit line at each redshift is shown in Fig.~\ref{fig:sfr_beta} and the values of the slope along with its uncertainty are listed in Table~\ref{tab:sfr_beta_table}. As seen from Table~\ref{tab:sfr_beta_table}, we find a $5\sigma$ significance dependence between $\beta$ and SFR at $z=6$, a $>5\sigma$ significance at $z=7$ and a $>3\sigma$ significance at $z=8$ and $z=9$. We also check if the choice of IMFs has any effect on the observed correlation between SFR and $\beta$ by re-deriving our SFR values with a Salpeter IMF at $z=6$. As explained in Section~\ref{sec:betamass}, the estimated SFRs are higher by 0.24 dex and this changes the slope from $0.20\pm0.04$ to $0.17\pm0.04$ at $z=6$.

Furthermore, to test whether our SFR to $\beta$ correlation is real and not a result of our applied dust correction to the SFRs, in addition to the dust corrected SFRs obtained from \citet{Kennicutt1998} and \citet{Meurer1999} relation, we also use the SFRs obtained from our SED fitting code (See Section~\ref{sec:properties}) to see if it has any effect on the observed $\beta$ and SFR relation. We find that the estimates obtained from SED fitting code and \citet{Kennicutt1998} equation are in good agreement for all galaxies, apart from a few galaxies with very high SFRs (SFR $>100\mathrm{M_{\odot}yr^{-1}}$). As seen from Fig.~\ref{fig:sfr_beta}, we do not find any galaxies with SFR $>100\mathrm{M_{\odot}yr^{-1}}$ and therefore we conclude that our observed SFR to $\beta$ correlation is not affected by the choice methods used for estimating SFRs.

Additionally, we investigate if any relation exists between $\beta$ and specific star formation rate (sSFR = SFR/M$_{*}$) by plotting $\beta$ as a function of sSFR. However, no trend is observed at $z=6,7,8$ and $9$, irrespective of the choice of IMFs. This suggests that whatever is setting $\beta$ is not a local process but a global one. This might be due to feedback or the halo retaining gas and dust, which otherwise would be ejected by supernova. These features make it unlikely that a variation of the IMF is what is producing the $\beta$ values we calculate. Star formation is in itself a local process and the range and the types of stars formed in a star formation event do not vary much if at all due to the property of the host galaxy. Because our trends correlate with the entire scale of the galaxy, namely its stellar mass, the processes determining the scale of $\beta$ must be due to features that scale with this, such as the total halo mass, or dust/metals and not the IMF.

Finally, as an additional check, we also conduct a principal component analysis (PCA) to see what the principal features are amongst $\beta$, $M_{1500}$, stellar mass and SFR. By carrying out a PCA analysis, we find that in the principal component space the variance is maximized along the principal component 1 (PC1), which explains 54$\%$ of the variance, and principal component 2 (PC2) explaining 26$\%$ of the variance. Inspecting the absolute values of the eigenvectors components in PC1, we observe that the principal features contributing to PC1 are stellar mass, SFR, $\beta$ and $M_{1500}$ with values of 0.61, 0.50, 0.45 and 0.40 respectively, confirming that stellar mass is the main principal feature.

\begin{deluxetable*}{cccccc}
\tabletypesize{\small}
\tablecaption{Median values of the UV spectral slope $\beta$ in different SFR bins}
\label{tab:sfr_beta_table}
\tablewidth{0pt}
\tablehead{
\colhead{$z$} & \colhead{Median $\beta$} & \colhead{Median $\beta$} & \colhead{Median $\beta$} & \colhead{Median $\beta$} & \colhead{$\beta$--SFR}\\[-0.1cm]
\colhead{$ $} & \colhead{$-2.0<\log\mathrm{SFR}<$} & \colhead{$-1.0<\log\mathrm{SFR}<$} &  \colhead{$0.0<\log\mathrm{SFR}<$} & \colhead{$1.0<\log\mathrm{SFR}<$} & \colhead{Slope}\\[-0.1cm]
\colhead{$ $} & \colhead{$-1.0$} & \colhead{$0.0$} & \colhead{$1.0$} & \colhead{$2.0$} & \colhead{$ $}
}
\startdata
	 6  & $-2.55_{-0.16}^{+0.13}$ & $-2.31_{-0.18}^{+0.10}$ & $-2.15_{-0.19}^{+0.15}$ & $-2.11_{-0.25}^{+0.31}$ & $0.20\pm0.04$\\
		 7 & $-2.71_{-0.05}^{+0.15}$ & $-2.32_{-0.17}^{+0.17}$ & $-2.05_{-0.22}^{+0.22}$ & $-1.52_{-0.28}^{+0.32}$ &$0.39\pm0.06$\\
		 8  & $..$ & $-2.56_{-0.07}^{+0.22}$ & $-2.23_{-0.25}^{+0.28}$ &  $..$ & $1.09\pm0.36$\\
		 9  & $..$ & $..$ & $-2.57_{-0.15}^{+0.30}$ & $-1.91_{-0.18}^{+0.34}$ & $0.85\pm0.17$ \\
\enddata
%\tablecomments{}
\end{deluxetable*}

\subsection{The main sequence correlation of stellar mass with SFR}
\begin{figure*}
\centering
\begin{minipage}{0.45\textwidth}
\centering
\includegraphics[width=1\textwidth, height=0.3\textheight]{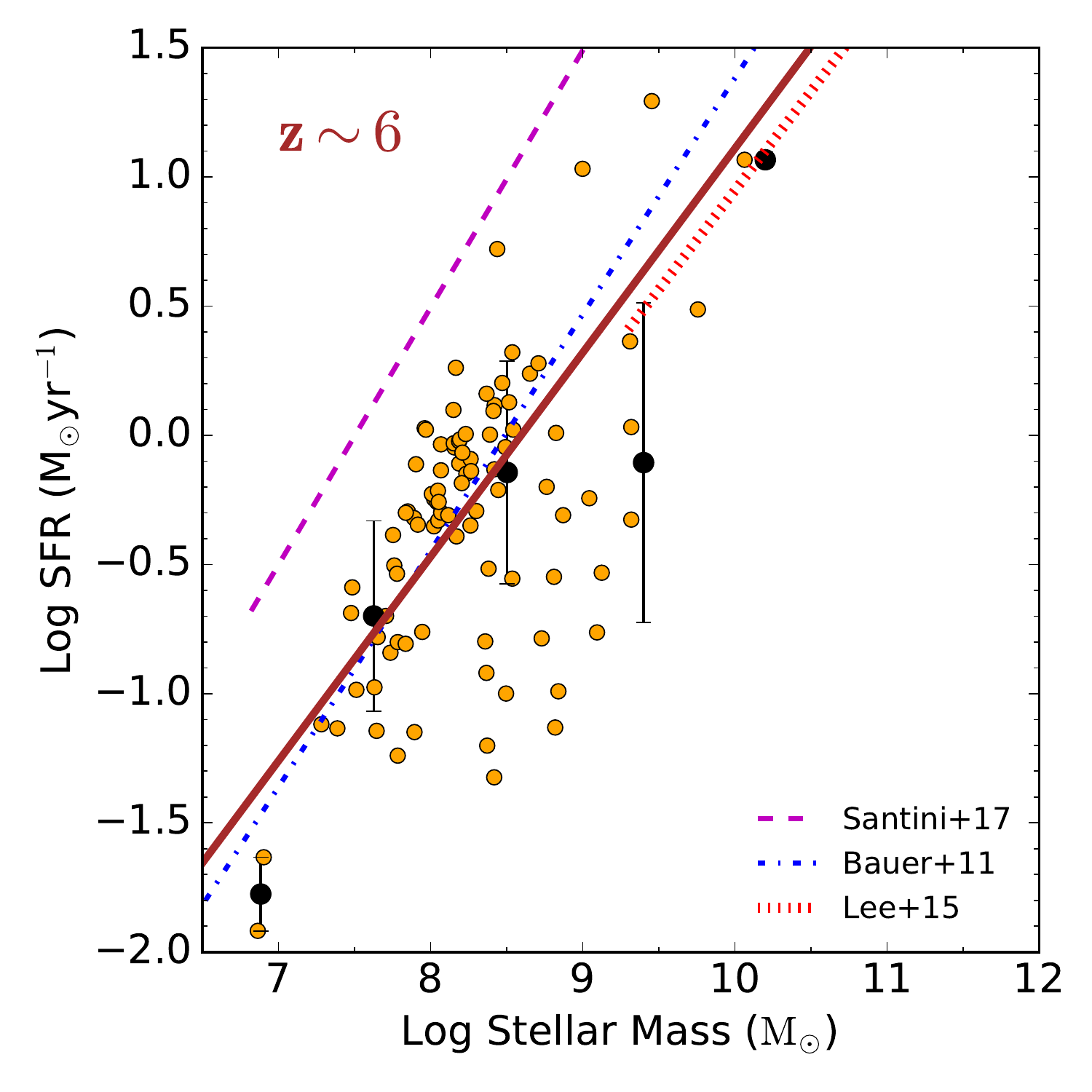}
\end{minipage}
\begin{minipage}{0.45\textwidth}
\centering
\includegraphics[width=1\textwidth, height=0.3\textheight]{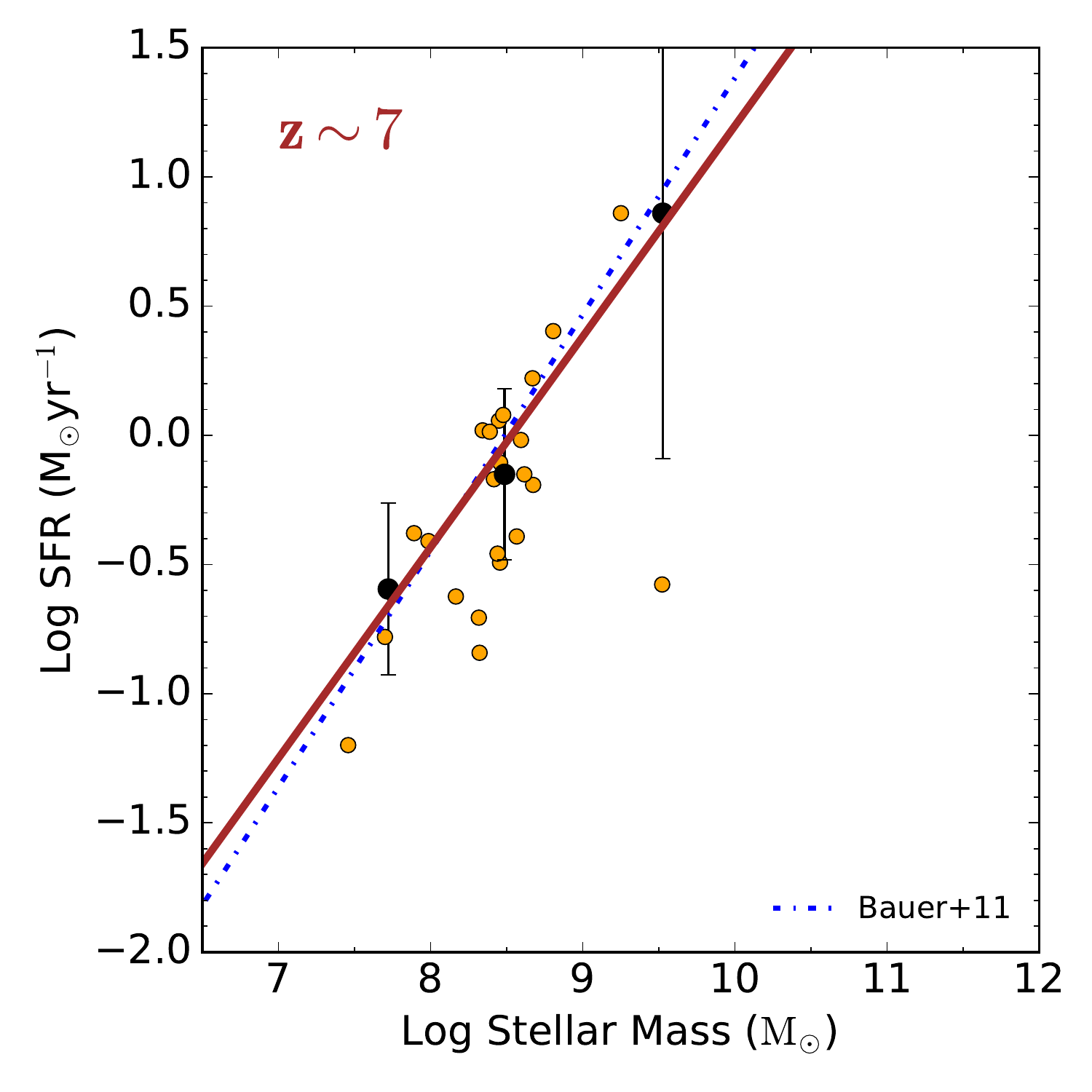}
\end{minipage}
\begin{minipage}{0.45\textwidth}
\centering
\includegraphics[width=1\textwidth, height=0.3\textheight]{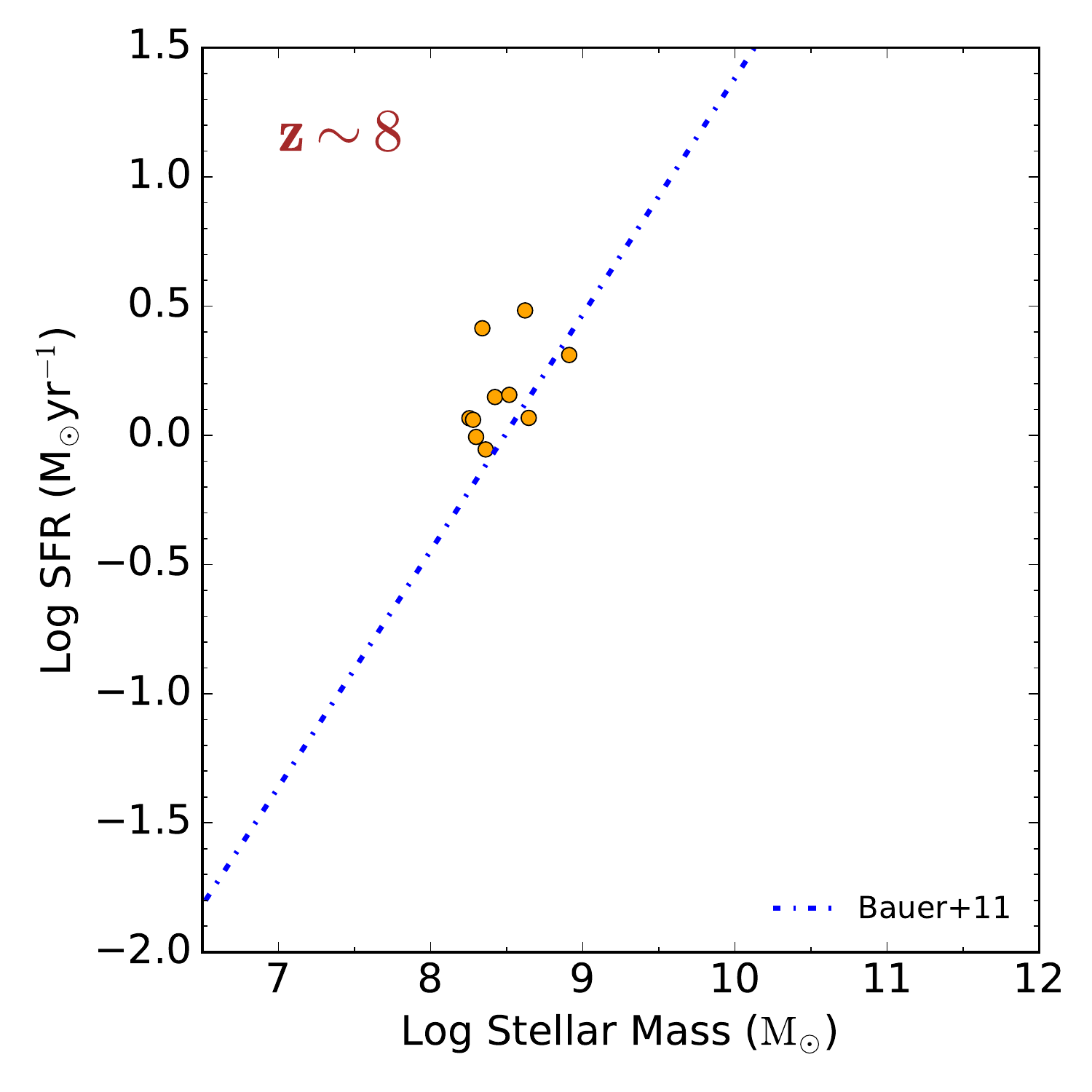}
\end{minipage}
\begin{minipage}{0.45\textwidth}
\centering
\includegraphics[width=1\textwidth, height=0.3\textheight]{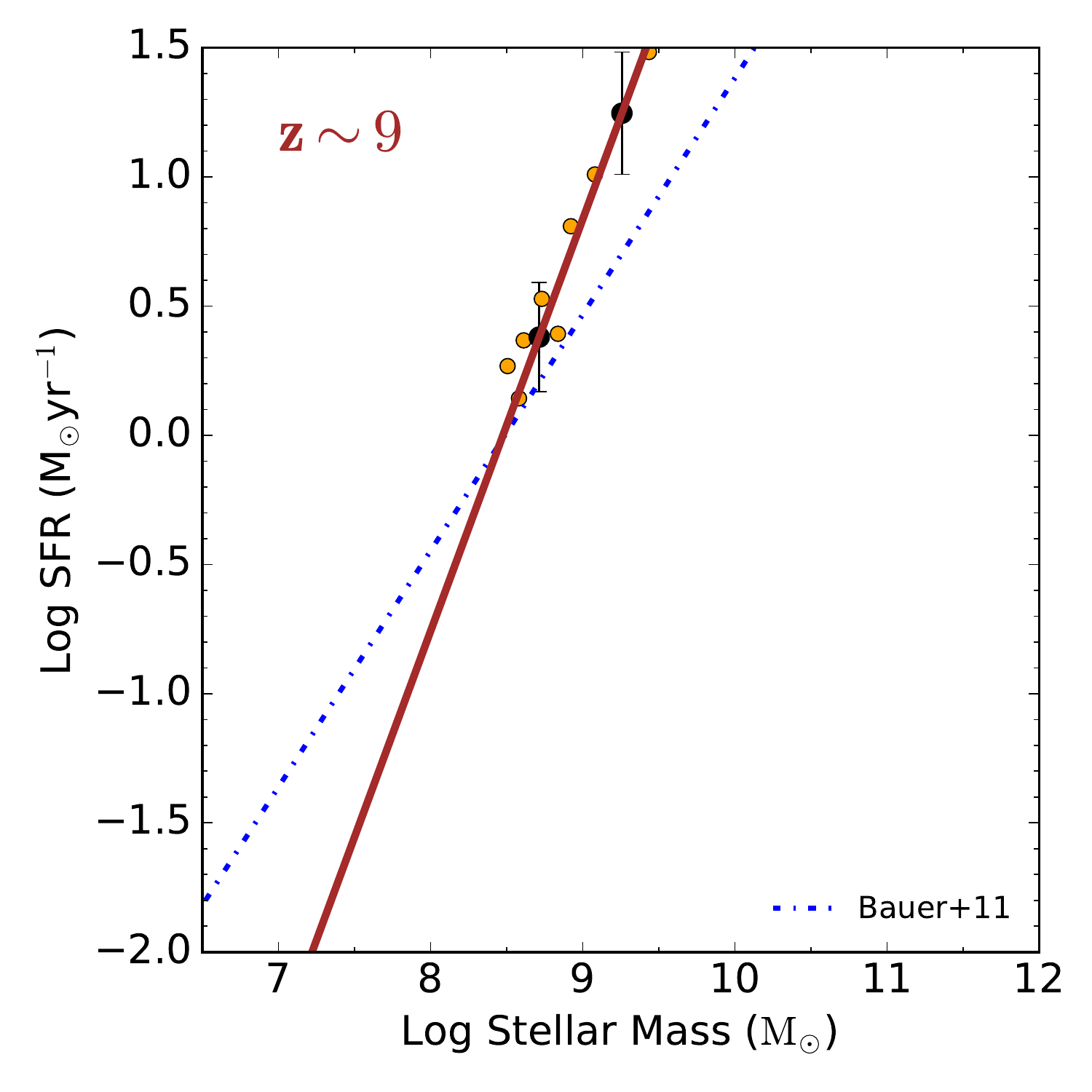}
\end{minipage}
\caption{The main sequence relation between stellar mass and SFR $z=6-9$. The filled yellow circles show the results for individual galaxies whereas the black circles show the median values of SFRs in each stellar mass bin of $1M_{\odot}$, with the uncertainties being the errors on the median. The solid red lines show a linear fit through the median points. The main sequence relation at $1.0<z<1.3$, $2.5<z<3$, and $5\leq z<6$ from \citet{Lee2015}, \citet{Bauer2011} and \citet{Santini2017} are shown by dotted red, dot-dashed blue and dashed magenta lines respectively at $z\sim6$. The dot-dashed blue line of \citet{Bauer2011} has been extrapolated to lower masses at $z\sim6$ and is copied on other redshifts as a reference point to examine the evolution in the main sequence relation. }
\label{fig:sfr_mass}
\end{figure*}

In this section we examine the main sequence relation between stellar mass and SFR to investigate whether any correlation exists between these quantities. To do this, we once again make use of the dust corrected demagnified star formation rates and demagnified stellar masses from \citet{Bhatawdekar2019}. In Fig.~\ref{fig:sfr_mass} we present SFR as a function of stellar mass at $z=6$ and for the first time at $z=7,8$ and $9$. Although the dynamic range in stellar mass is dramatically reduced at $z=8$ and $z=9$, it is evident that an overall trend for rising SFRs with increasing stellar mass exists at these highest redshifts.

To quantify this, we compute the median values of SFRs in different mass bins of $6<\mathrm{log}M/M_{\odot}<7$, $7<\mathrm{log}M/M_{\odot}<8$, $8<\mathrm{log}M/M_{\odot}<9$, $9<\mathrm{log}M/M_{\odot}<10$ and $10<\mathrm{log}M/M_{\odot}<11$ and fit a first-order polynomial through the median data points at each redshift, revealing a slope of $0.76\pm0.16$ ($\sim5\sigma$ significance), $0.81\pm0.19$ ($\sim4\sigma$ significance) and $1.59\pm0.37$ ($\sim4\sigma$ significance) at $z=6,7$ and $9$ respectively. In Fig.~\ref{fig:sfr_mass}, we show the best-fit line at each redshift, except $z=8$ since there is only one median data point. Additionally, at $z=6$ we test if the choice of IMFs has any effect on the observed correlation and find that with a Salpeter IMF the slope changes from $0.76\pm0.16$ to $0.64\pm0.17$ at this redshift.

We also show in Fig.~\ref{fig:sfr_mass} the main sequence relation at $1.0<z<1.3$, $2.5<z<3$, and $5\leq z<6$ from \citet{Lee2015}, \citet{Bauer2011} and \citet{Santini2017} respectively for comparison. As seen, our main sequence relation is in agreement with \citet{Lee2015} and \citet{Bauer2011} at $z\sim6$ but not with \citet{Santini2017}. Considering that \citet{Santini2017} estimate their SFRs with the new \citet{Kennicutt2012} factor, we recalculate our SFRs with this relation to see if this has any effect on the main sequence comparison. As demonstrated in \citet{Kennicutt2012}, we confirm that the ratio of SFRs derived with the new equation to the old \citet{Kennicutt1998} equation is 0.63. These lower SFRs are a result of updated SSPs and a different IMF, however, the trend for rising SFRs with increasing stellar mass is unaffected.

Similarly, to examine if there is any evolution in the main sequence relation, we extrapolate the \citet{Bauer2011} relation to lower masses and copy it on to the other redshifts as a reference point. It is clear from Fig.~\ref{fig:sfr_mass} that there is no evolution in the main sequence relation from $z=6-9$.

We derive the stellar masses of our sample of galaxies with our SED fitting code (See \citet{Bhatawdekar2019} for more details), which for objects at these redshifts in most cases will likely fit star-formation histories that are effectively constant star-formation rates. Similarly, the calibration between UV luminosity and star-formation rates derived by \citet{Kennicutt1998} equation is also fundamentally based upon the assumption of a constant star-formation history. Therefore, it is important to investigate if our observed correlation between stellar mass and SFR is real. Thus, to examine if our choice of SFR estimation method is affecting the observed stellar mass to SFR relation, we also use the SFRs obtained from our SED fitting code. As stated in Section~\ref{sec:betasfr}, the estimates obtained from our SED fitting code and the \citet{Kennicutt1998} equation agree well, with the exception of a few galaxies with very high SFRs (SFR $>100\mathrm{M_{\odot}yr^{-1}}$). As we are not finding any galaxies with SFR $>100\mathrm{M_{\odot}yr^{-1}}$, we conclude that our stellar mass to SFR relation remains unaffected irrespective of the choice of methods.

\subsection{Physical meaning of $\beta$}

There are a number of factors that influence the rest frame UV colours. The UV continuum  produced by young and massive but short-lived O and B stars is dependent on the surface temperature, mass and metallicity of stars. This means that the stellar population continuum will also be dependent on the distribution of the masses (determined by the IMF and SFH) and metallicities. The nebular emission also has an impact on the UV continuum such that a very blue slope suggests that the UV light is not significantly contaminated by redder nebular continuum (\citealt{Dunlop2012, Robertson2010}). Finally, the UV continuum is also affected by the distribution of dust with respect to the galaxies.

For example, since the UV continuum is dependent on its initial mass function, with massive stars producing bluer slopes, the IMF of a stellar population can likely affect the slope of a composite stellar population. Thus, a top-heavy IMF (with a larger number of massive stars) will produce bluer values of $\beta$.

The UV continuum slope is also affected by the star formation history such that with extended periods of star formation the massive stars will evolve quickly from the main sequence resulting in a redder slope, while continuous bursts of shorter duration will produce a bluer slope. 

Similarly, the overall metallicity will also affect the UV colours such that a higher metallicity star will generate reduced energy, resulting in a redder slope and vice-versa.

Finally, since the reddening (caused by extinction due to dust grains) of an object is inversely proportional to the wavelength of optical light, the UV continuum is most affected by dust. The shape of the dust attenuation curve, on the other hand, is sensitive to the source-dust geometry, grain size distribution etc. Due to these factors, estimating dust attenuation in external galaxies is much more complicated. Since the ``Calzetti attenuation law'' \citep{Calzetti1994} is derived from a sample of nearby starburst galaxies, to what extent it is applicable to other systems at high redshift is debatable.

All these factors, the IMF, SFH, metallicity and dust, affect the UV slope $\beta$ of stellar populations. Establishing the effects of these factors is challenging, particularly at high redshift. With the increased spectral coverage beyond 1.6$\mu$m, JWST will provide better constraints on the UV slope $\beta$ for objects at $z>8$.

\section{SUMMARY}\label{sec:summary}
In this paper we investigate the UV spectral slope $\beta$ for a sample of high redshift galaxies we previously located at $z=6-9$ in the MACSJ0416 cluster and its parallel field. With this study, we offer insight into the rest-frame UV colours of galaxies in a wide magnitude range of $-22<\mathrm{M_{UV}<-13}$ within the HFF dataset.

We utilize the galaxy samples at $z=6-9$ from \citet{Bhatawdekar2019} and measure the value of UV spectral slope $\beta$ by fitting a power law to the best-fit model spectrum in the windows defined by \citet{Calzetti1994}. With this, we derive the rest-frame UV colours of galaxies out to $z=9$ within the HFF program probing magnitudes as faint as $M\mathrm{_{UV}=-13.5}$ at $z=6$. We also show with the help of simulations that this method is quite effective at recovering the correct values of $\beta$ with a small scatter. We measure the median value of $\beta$ in each redshift bin, as well as in separate bins split by UV luminosity, stellar mass and star formation rates for a more reliable result. Our key conclusions are as follows:
\begin{enumerate}[itemsep=0pt,parsep=0pt]
 \item We find no significant correlation between $\beta$ and rest-frame UV magnitude $M_{1500}$ at all redshifts probed in this work. However, some evidence for a mild evolution of the median $\beta$ values (from $\beta=-2.22_{-0.12}^{+0.08}$ at $z\sim6$ to $\beta=-2.52_{-0.20}^{+0.32}$ at $z\sim9$) for galaxies at all luminosities from $z=6-9$ is observed, presumably due to rising dust extinction. The average value of $\beta$, however, becomes redder at a upper luminosity cut of $M_{\mathrm{UV}}=-18$, suggesting that faint galaxies in our sample are likely causing this apparent evolution. 
 \item At $z=7$, the bluest median value of our sample is $\beta=-2.32_{-0.23}^{+0.30}$, which is redder than previously reported values at this redshift in the literature. Similarly, with the help of our SED fitting method, we find that our bluest data point has a median value of $\beta=-2.63_{-0.43}^{+0.52}$ at $z\sim9$, implying no evidence as yet for extreme stellar populations at $z>6$ with HFF.
 \item Fitting for a linear correlation, we find a strong correlation between $\beta$ and stellar mass, such that lower mass galaxies exhibit bluer UV slopes. It also appears that low mass galaxies at $\log M/M_{\odot}<9$ become bluer with increasing redshift, whereas the massive galaxies at $\log M/M_{\odot}>9$ appear to exhibit a nearly constant $\beta$ at each redshift. 
 \item We investigate the correlation between $\beta$ and SFR and find that there is a strong correlation, such that galaxies with low SFRs exhibit bluer slopes.
 \item Examining the relation between $\beta$ and sSFR, we observe no trend between these quantities at $z=6,7,8$ and $9$, suggesting that whatever is setting $\beta$ is not a local process but a global one.
 \item Finally, we investigate the main sequence relationship between stellar mass and SFR and find an overall trend for rising SFRs with increasing stellar mass at $z=6-9$. However, more data are needed to confirm the trend at the highest redshifts.
\end{enumerate}

All these results suggest that even with the deepest HST imaging possible, combined with the power of gravitational lensing, we are still not reaching the first stars and galaxies at $z\sim9$. While the sample size of this study is small and future studies including the complete HFF dataset will shed further light on these issues at the highest redshifts we probe in this study, it is clear that galaxy and structure formation predates even this very early redshift. JWST will certainly provide a clearer picture of this when it examines galaxies at even higher redshifts where ultimately Pop III stellar populations will be discovered.

\section{ACKNOWLEDGEMENTS}
The authors thank the referee for their careful reading of the paper and valuable comments, which has greatly improved this paper. This work is based on the observations made with the NASA/ESA Hubble Space Telescope, obtained from the Mikulski Archive for Space Telescopes (MAST) at the Space Telescope Science Institute (STScI), which is operated by the Association of Universities for Research in Astronomy, Inc., under NASA contract NAS 5-26555. This work is also based on observations made with the Spitzer Space Telescope, which is operated by the Jet Propulsion Laboratory (JPL), California Institute of Technology under a contract with NASA and uses the data taken with the Hawk-I instrument on the European Southern Observatory (ESO) Very Large telescope (VLT) from ESO programme 092.A-0472. This work utilizes gravitational lensing models produced by PIs Brada\u{c}, Ebeling, Merten \& Zitrin, Sharon, and Williams funded as part of the \textit{HST} Frontier Fields program conducted by STScI. The lens models were obtained from the MAST. R.A.B gratefully acknowledges support from the European Space Agency (ESA) Research Fellowship.
\software{GALAPAGOS \citep{Barden2012}, GALFIT \citep{Peng2002}, T-PHOT code \citep{Merlin2015}, EAZY \citep{Brammer2008}, SMpy \citep{Duncan2014}}.

\bibliography{hff_uvslopes}{}
\bibliographystyle{aasjournal}

\end{document}